\documentclass[%
 aip,
 jmp,%
 amsmath,amssymb,
 reprint,%
]{revtex4-2}

\usepackage{graphicx}
\usepackage{dcolumn}
\usepackage{bm}
\newcommand*{\plimsoll}{{\ensuremath{-\kern-4pt{\ominus}\kern-4pt-}}}
\usepackage{color}

\usepackage{graphicx}
\usepackage{dcolumn}
\usepackage{bm}
\usepackage{cancel}
\usepackage{graphics}

\begin{document}

\preprint{APS/123-QED}

\title{Edge-effects dominate copying thermodynamics for finite-length molecular oligomers}
\author{Jenny M. Poulton}
\author{Thomas E. Ouldridge}%
 \email{t.ouldridge@imperial.ac.uk}
\affiliation{%
Departmant of Bioengineering and Centre for Synthetic Biology, Imperial College London, London, SW7 2AZ, UK.
}%

\date{\today}

\begin{abstract}
A signature feature of living systems is the ability to produce copies of information-carrying  molecular templates such as DNA. These copies are made by assembling a set of monomer molecules into a linear macromolecule with a sequence determined by the template. The copies produced are finite-length ``oligomers", or short polymers, which must detach from their template in the long term. We explore the role of the resultant initiation and termination of the copy process in the thermodynamics of copying. By splitting the free-energy change of copy formation into informational and chemical terms, we show that, surprisingly, copy accuracy plays no direct role in the overall thermodynamics. Instead, finite-length templates function as highly-selective engines that interconvert chemical and information-based free energy stored in the environment; it is thermodynamically costly to produce outputs that are more similar to the  oligomers in the environment  than sequences obtained by randomly sampling monomers. In contrast to previous work\cite{Poulton}, any excess free energy stored in correlations between copy and template sequences is lost when the copy fully detaches and mixes with the environment; these correlations therefore do not feature in the overall thermodynamics. Previously-derived constraints on copy accuracy therefore only manifest as kinetic barriers experienced while the copy is template attached; these barriers are easily surmounted by shorter oligomers. 

\end{abstract}
\maketitle

\begin{figure*}
    \centering
    \includegraphics[scale=0.3]{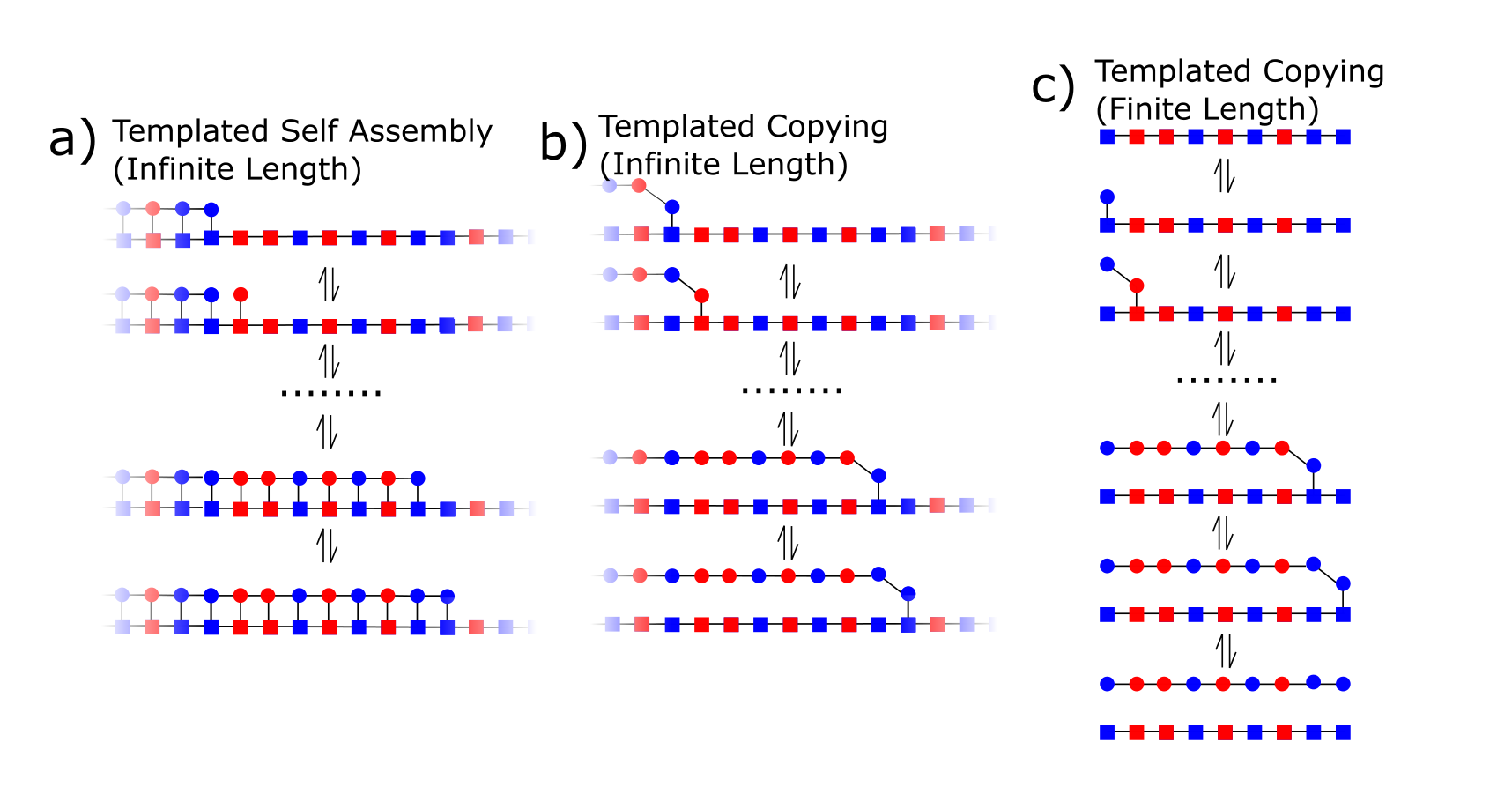}
    \caption{{\it Distinct mechanisms for producing sequence-matched assemblies from a template.} (a) A model for templated self assembly on an infinitely long template, with a two-monomer alphabet for both template and product.  Here favourable bonds retained between copy and template stabilise the low entropy product state. (b) A model of true templated copying on an infinitely-long template. The nascent copy detaches continuously from behind the leading edge of polymer growth, but the process never reaches the final monomer in the template and therefore the product remains bound by a single site. (c)  A model of persistent copying on a finite oligomer template, including full dissociation of full length oligomers and single monomers into surrounding baths.}
    \label{Diagrams}
\end{figure*}

\begin{figure}
    \centering
    \includegraphics[scale=0.25]{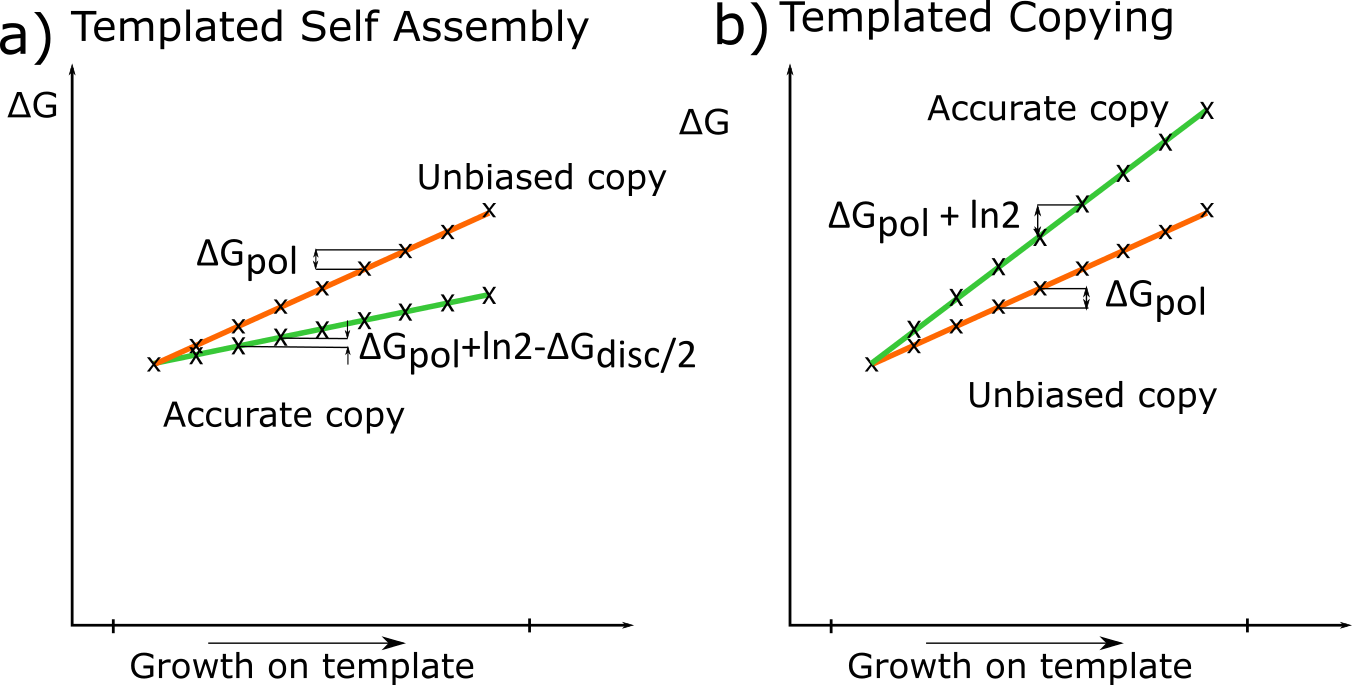}
    \caption{{\it Thermodynamics of templated self-assembly and copying for infinite-length templates.} Free-energy profiles of (a) templated self-assembly on an infinite length template and (b) templated copying with continuous separation.  templated copying with continuous separation on an infinite-length template. \(\Delta G_{\rm pol}\) is the total free energy change associated with adding a monomer to the growing polymer in an unbiased fashion and sets the slope of the orange line. In templated self-assembly, a perfectly accurate copy can be more be more favourable (have a more negative gradient) if the discrimination free energy difference is strong enough. For sufficiently large \(\Delta G_{\rm disc}\) the green line can be thermodynamically down hill. By contrast, for templated copying on an infinite template, accuracy increases the gradient because the low entropy of the product cannot be compensated.}
    \label{NewFig1}
\end{figure}

Information transfer is the essence of the central dogma of molecular biology. The cell has sophisticated biomachinary which directs the copying of information from DNA into RNA, and then from RNA into proteins; complex molecules which perform varied roles within an organism\cite{Alberts2002}.  The folding of proteins into their functional form is encoded directly into their sequence, meaning that accurate copying from DNA, via RNA, to proteins is vital for the functioning of an organism. In humans, there are around 20,000 functional proteins (peptide polymers) made from a library of 20 amino acids (monomer units), and the targeted assembly of these specific sequences by the cell is an astonishing accomplishment given the combinatorial space available to the system. The scale of this accomplishment can be highlighted by considering the difficulty of assembling such structures without templates.

Without a template, the cell would have to self-assemble proteins using hard-coded interactions between amino acid monomers targeting a specific structure as a free-energy minimum. This task is made difficult by the existence of many unintended, competing structures in cases where monomers can be used in multiple different structures\cite{murugan2015multifarious}]. As a result, even simple structures require many distinct monomer types to self-assemble reliably from their constituents. For example, a simple 4-armed cross, with arms of length two surrounding a central monomer, requires three distinct types of monomer to assemble reliably\cite{johnston2011evolutionary}. 

Targeting the reliable self-assembly of a single protein structure, containing hundreds of amino acid residues, would therefore be an extraordinary challenge with only 20 distinct monomers. Cells, however, need to target the assembly of tens of thousands of distinct proteins with the same 20 building blocks -- it is simply impossible to hard-code this level of complexity into the interactions between the amino acids themselves\cite{murugan2015multifarious}. 

Information transfer via templating solves this problem, because an arbitrary sequence of amino acids can be assembled in a way that does not depend on the interactions between the amino acids themselves. These interactions are only subsequently used to fold the protein into a specific secondary and tertiary structure, given the information in the sequence. It should be noted, however, that templating only solves this problem if the template is reusable. Without this crucial property, we would have merely pushed the problem back a layer; a separate template would have to be assembled for each product created, recreating the problem solved by templating. Thus copy and template must separate.

Numerous groups have worked on developing synthetic analogs of the processes of the central dogma. A major challenge is product inhibition, the tendency for copies to remain bound to their template, due to the cooperative interaction of the monomers in the copy with the template\cite{Vidonne, orgel, colomb2015}. Cooperativity grows with length, meaning product inhibition is a more difficult problem to overcome for longer polymers. Consequently, progress in building synthetic copying systems where copy and template spontaneously separate -- as in the processes of the central dogma --  has been limited. The most successful examples of copying rely on 
non-chemical or time-varying conditions to drive separation; the copy first forms on the template and then separates via mechanical scission ~\cite{Schulman,Schulman2}, heat ~\cite{braun2004}, or a change in chemical conditions ~\cite{zhuo2019litters}. Due to the challenge of separating products from templates, synthetic examples of the biologically-relevant context in which copying is chemically-driven and autonomous have only involved dimers and trimers ~\cite{Sievers,Lincoln}.

This difficulty in recreating a fundamental biological phenomenon in a minimal synthetic context suggests a gap in our basic understanding. Indeed, previous theoretical work has omitted the separation of copy and template \cite{Bennett,Cady,Andrieux,Sartori1,Sartori2,esposito2010,EHRENBERG1980333,Johansson}, instead studying templated self assembly, as illustrated in Fig. \ref{Diagrams}(a).  In previous work \cite{Poulton}, we considered the system shown in fig. \ref{Diagrams}(b). Here, a growing polymer separates sequentially from the template as it grows, analogously to a nascent polypeptide chain or RNA strand during transcription or translation, respectively. It does not, therefore, build up a stronger and stronger attachment to the template. Before describing our results, we summarise the fundamental thermodynamics of these two models.

For both the processes in fig.~\ref{Diagrams}\,(a) and (b), producing a sequence that matches the template corresponds to producing a low entropy product. For a system with two monomer types, the reduction in sequence entropy relative to an unbiased random sequence is $\Delta H = \ln 2$ per monomer (throughout this article, we represent entropies in units of $k_B$ and energies in units of $k_BT$).

 To understand the role of this entropy difference, and the thermodynamic distinction between templated copying and self-assembly, consider the plots in fig.~\ref{NewFig1}. These plots give a schematic representation of the free-energy profile of copy formation -- for simplicity, we consider models in which growth is ``tightly coupled", without kinetic proofreading cycles\cite{Hopfield}, and in which the monomer interactions are symmetric (all complementary interactions are equivalent, as are all non-complementary interactions). We define $\Delta G_{\rm pol}$ as the gradient of the free-energy profile for the production of an unbiased, random sequence. These unbiased sequences will tend to grow if the slope is negative, $\Delta G_{\rm pol}<0$. For a perfectly accurate copy, the reduction in entropy of $\Delta H =  \ln 2$ per monomer relative to a random sequence will tend to make the gradient of the free-energy profile more positive by $\ln 2$; the entropic cost thus tends to inhibit growth relative to an unbiased random sequence. 
 
  In a model that omits copy/template separation entirely (fig. \ref{Diagrams}(a)), the stronger copy/template interaction of a correctly-matched copy sequence can more than compensate for this reduction in entropy, as illustrated in fig. \ref{NewFig1}(a). Let \(\Delta  G_{\rm disc}>0\) be the free-energy difference between a mismatched monomer binding to the template and a correctly-matched monomer binding to the template. A correctly-matched copy will have its free energy reduced by  \(\Delta  G_{\rm disc}/2\) per monomer relative to an unbiased random sequence. For \(\Delta  G_{\rm disc}/2>\ln2\), the perfectly accurate copy is thermodynamically more favourable than an unbiased sequence, having a more negative slope in the free-energy profile and growing at a less negative $\Delta G_{\rm pol}$. In principle, \(\Delta G_{\rm disc}\) could be arbitrarily large, making the perfectly-matched sequence arbitrarily thermodynamically favoured relative  to alternatives. Polymers produced by templated self-assembly can thus be arbitrarily accurate in thermodynamic equilibrium

  Explicitly considering the disruption of copy-template bonds, as in fig.~\ref{Diagrams}\,(b) changes this analysis \cite{Ouldridge,Poulton}. Since all copy-template bonds are transient, there is no long-term thermodynamic benefit to incorporating a correctly-matched monomer into the copy. Consequently, the entropic cost of creating an accurate copy cannot be overcome by $\Delta G_{\rm disc}$ terms, and the free-energy profile of an accurate copy always has a steeper slope than a random sequence (fig.~\ref{NewFig1}(b)). Unlike templated self-assembly, therefore, there is necessarily a thermodynamic cost to accuracy for templated copying of an infinite-length polymer: a more negative $\Delta G_{\rm pol}$ (a stronger thermodynamic drive towards polymerisation) is required to grow an accurate copy. In addition, the higher free energy of the accurate sequence means that, unlike in templated self-assembly, accurate copying necessitates generating a far-from-equilibrium product. Mechanistically, producing a specific non-equilibrium state is a very different challenge from allowing the system to relax to a stable equilibrium. Moreover, from the thermodynamic perspective, the template is playing a distinct role. It acts as an engine that transduces excess free energy stored in the input monomers into excess free energy stored in the low entropy product sequence, rather than as a reactant whose sequence directly contributes to the free energy of products. It was observed in \cite{Poulton}, that, for finite binding free energies,this transduction could never be 100\% efficient. Producing an accurate copy was therefore seen to be necessarily thermodynamically irreversible.

These previously-obtained results, both for templated self-assembly \cite{Bennett,Cady,Andrieux,Sartori1,Sartori2,esposito2010,EHRENBERG1980333,Johansson} and for a polymer that continuously separates from its template \cite{Poulton}, were derived for infinite-length polymers. The tip of the growing copy is assumed to reach a steady velocity along the template, and the identity of monomers measured relative to this tip reach a stationary distribution. Initiation and termination of the polymerization process were ignored, and even in Ref.~\cite{Poulton} the copy remains attached by a single bond - complete detachment was neglected.

This approximation might be reasonable for some long biopolymers {\it in vivo}, it is a poor approximation for the copying of shorter oligomers and dimers.  Given that synthetic systems are currently limited to short oligomers, and that early life is likely to have created short oligomers before the origin of complex enzyme-based copying machinery, it is worth studying initiation and termination in more detail. Equally, ``charged" tRNA -  hybrid molecules of one tRNA codon and its matching amino acid that are necessary for the process of RNA translation into proteins - are dimerised via a specific template enzyme called a synthetase~\cite{gomez2020aminoacyl}.  This process is essentially the copying of a dimer template, and so the creation of short copies is highly relevant even to extant biology.


In this article we probe the consequences of the ``edge-effects" of initial attachment and final detachment on the copying of oligomer sequences.
We first consider the free-energy change for the production of a single finite-length copy under constant external conditions, separating it into chemical and informational terms.  Using  dimerisation as an example,  we show that the overall thermodynamic constraints on information transfer are fundamentally altered relative to infinite-length polymers: in general there is no entropic cost to accurately reproducing the template sequence, in direct contrast to previous work on infinite length templates\cite{Poulton,Bennett,Cady,Andrieux,Sartori1,Sartori2,esposito2010,EHRENBERG1980333,Johansson}. As in Ref.~\cite{Poulton}, the template acts as an information engine that doesn't directly contribute to the relative free energies of products, but here its role is to selectively couple to a subset of a large number of out-of-equilibrium molecular reservoirs; the relationship between the copy sequence and these reservoirs sets the overall thermodynamics. Although these results hold for polymers of arbitrary finite length, we nonetheless observe a gradual cross-over to the previously predicted constraints on accuracy  \cite{Poulton} when studying a particular dynamical model of copying with longer polymers (Fig.~\ref{NewFig1}\,(c)). The thermodynamic constraints on accuracy in the infinite length limit \cite{Poulton} instead become kinetic barriers for finite length templates.


%
\begin{figure}
    \includegraphics[scale=0.47]{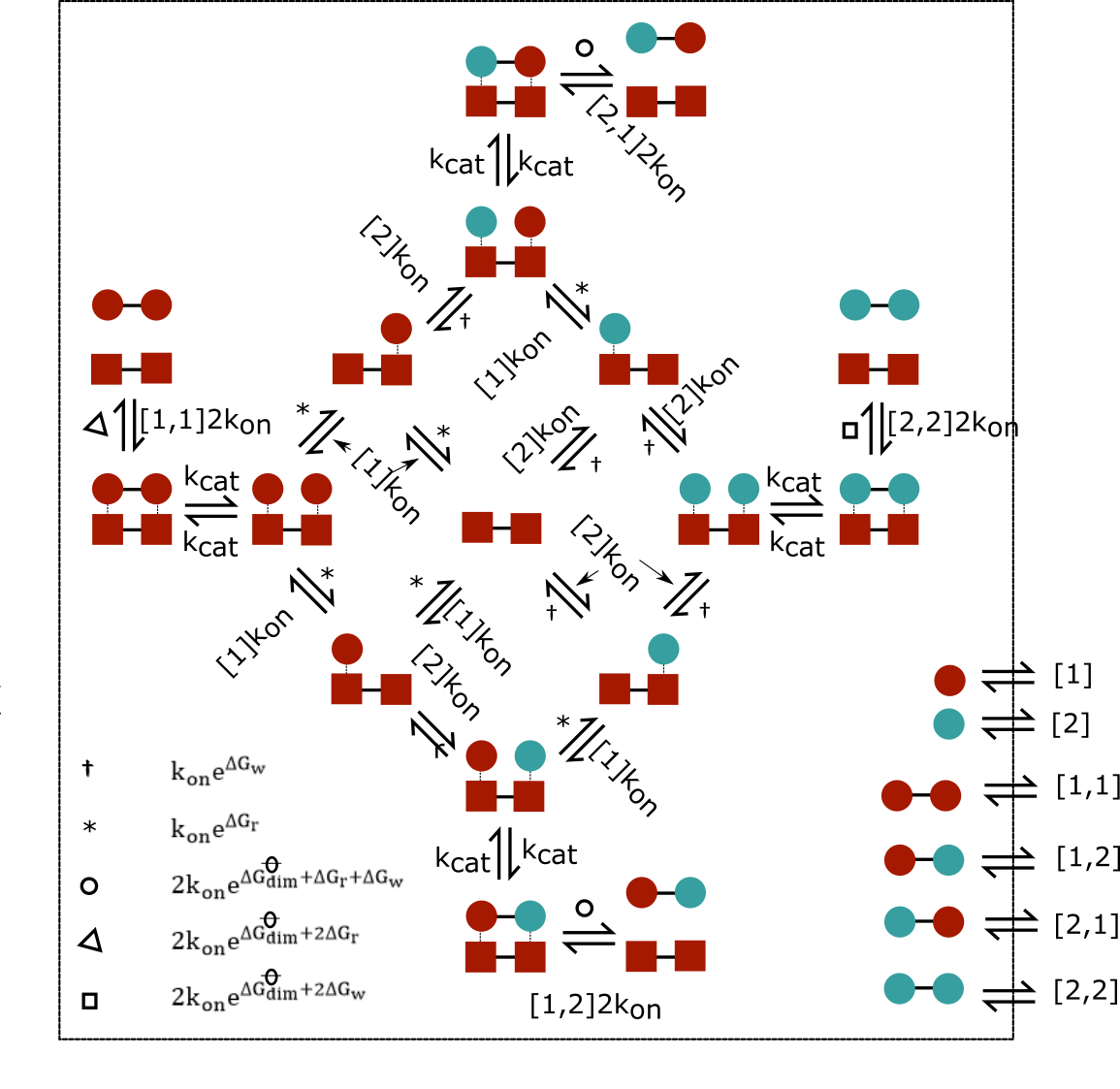}
    \caption{{\it Minimal model of oligomerisation illustrated with dimers.} A template (squares) interacts with baths of monomers and dimers of a second class of molecules (circles) which is present as type ``1" or type ``2". Monomers can bind to the template and dimerize, while dimers binding to the template can be destroyed or interconverted. 
    The standard dimerisation free-energy is \(\Delta G^\plimsoll_{\rm dim}\) for all sequences, but matching and non-matching varieties of monomer bind to the template with strengths \(\Delta G_r\) and \(\Delta G_w\), respectively,  allowing selectivity. Rate constants $k_{\rm cat}$ and $k_{\rm on}$ define the dynamics.}
    \label{Fig1}
\end{figure}

\section{Model of oligomerisation} 
\label{sec:model}
Fig.~\ref{Fig1} shows a dimerisation system, a prototype for a broader class of oligomerisation systems. A solvated template dimer carries information in its sequence of monomer units; in this case, the sequence is 1,1. The template is coupled to large baths of monomers and oligomers (in this case dimers) of a distinct second type of molecule, like a DNA template in a bath of RNA nucleotides and oligomers. This second type of molecule also  comes in multiple varieties -- in this case two -- and can interact with the template in a sequence-specific way. In  Fig.~\ref{Fig1}, we propose a particular thermodynamically-consistent model for dimer production (for simplicity, we do not allow for kinetic proofreading cycles \cite{Hopfield} in our analysis). Dimers from the baths can also be broken down into their component monomers by the template.

The parameters in Fig.~\ref{Fig1} can be divided into two classes; those that are properties of the copy species in isolation, and those that relate to the interaction of the copy species with the template. With regard to the former, we define \(\Delta G^{\plimsoll}_{\rm dim}\) as the free-energy change of dimerisation at the reference concentration for all sequences. We assume \(\Delta G^{\plimsoll}_{\rm dim}\) is sequence-independent, since arbitrary copy sequences must be producible given the right template. The concentrations of the two monomer types  are \([1]\) and \([2]\) (defined with respect to a standard reference concentration), and the four polymer types are \([1,1]\), \([1,2]\),\([2,1]\) and \([2,2]\).

The properties of the copy species alone will set the thermodynamic constraints on the system derived in section~\ref{sec:TD}. The details of the interaction between copy molecules and template will determine how a specific example system behaves with respect to these constraints. For the dimerisation example in Fig.~\ref{Fig1}, \(\Delta G^\plimsoll_{r} \) is the standard free-energy change of a matching monomer binding to the template; \(\Delta G^\plimsoll_{w}\), is the equivalent for a mismatch. The difference between the two, \(\Delta  G_{\rm disc}=\Delta G_{w}^\plimsoll-\Delta G^\plimsoll_{r}\), is the discrimination free energy. Parameters $k_{\rm cat}$ and $k_{\rm on}$ set the absolute values of transition rates between states.

In the model of Fig.~\ref{Fig1}, we assume that any free energy released by dimerisation is used to destabilise the bonds between dimer and template. The state with two monomers and the states with a dimer bound to the template therefore have equal free energy, with \(k_{\rm cat}\) the underlying rate of the formation or breaking of the backbone bond. Such a free-energy landscape has been proposed to be optimal for minimising product inhibition~\cite{deshpande2020optimizing}. Here, matching and non-matching monomers bind to the template with the same rate constant \(k_{\rm on}\), but since \(\Delta G^\plimsoll_{r}<\Delta G^\plimsoll_{w}\), mismatches detach faster. Since we use an unbiased \(m( s)\) in all case studies, the symmetry of the problem gives identical physics for all template sequences; we shall use 1,1 for clarity.

 \section{Results}
\subsection{Thermodynamics of oligomerisation}
\label{sec:TD}
 For the dimerisation model, the total free-energy change of the baths upon creating a single dimer of sequence ${\bf s}$ is
\(
 \Delta G({\bf s})=\Delta G_{\rm{dim}}^\plimsoll+ \left(\ln{[{\bf s}]}-\ln{[s_{1}][s_{2}]}\right)
\);  \({\bf s}\) is the arbitrary sequence \(s_{1},s_{2}\).
Since the template itself acts catalytically \cite{Ouldridge}, its free-energy is unchanged.
Since we do not allow for kinetic proofreading cycles \cite{Hopfield} in our analysis, the process is tightly coupled and there is no unknown and variable number of fuel-consuming futile cycles. The free energy change of the dimerisation process is therefore unambiguously defined as \(\Delta G({\bf s})\).

This free-energy change of dimerisation can be generalised straight-forwardly to oligomers of length \(|{\bf{s}}|\), %
\begin{equation}
 \Delta G({\bf s})=(|{\bf s}|-1)\Delta G^\plimsoll_{\rm{dim}}+ \left(\ln{[{\bf s}]}-\ln{\prod_{i=1}^{|{\bf s}|}[s_{i}]}\right).
\end{equation}\\
where \(\Delta G^\plimsoll_{\rm{dim}}\) is now the standard free energy change for adding any monomer to the end of a copy oligomer in solution. 

Let \(J({\bf s})\) be the expected net rate at which sequence \({\bf s}\) is produced by the the system when the template reaches a steady state.
The average rate of change of free-energy is then \(\Delta \dot{G}=\sum_{\bf s} J({\bf s})\Delta G({\bf s})\). We define the normalised flux; \(q({\bf s})=J({\bf s})/J_{\rm{tot}}\). We further define the following probability distributions: \(p({\bf s})=[{\bf s}]/[S_{\rm{tot}}]\), the probability of picking an oligomer of sequence \(\bf{s}\) from the oligomers with total concentration \([S_{\rm{tot}}]\); \(m({s})=[{ s}]/[M_{\rm tot}]\), the probability of picking a monomer of type \({s}\) from the monomers with total concentration of \([M_{\rm tot}]\); and \(t({\bf s})= \prod_{i}m(s_{i})\), which corresponds to the probability of the sequence \({\bf s}\) occurring by selecting monomers randomly from the monomer pools. In these terms, 
\begin{align*}
&\Delta \dot{G}=J_{\rm{tot}}\left(\sum_{{\bf s}}q({\bf s})\left(|{\bf s}|-1\right)\Delta G^\plimsoll_{\rm{dim}}\right)\\
+&J_{\rm{tot}}\left( \sum_{{\bf s}}q({\bf s})\ln{\frac{p({\bf s})}{t(\bf s)}}+\sum_{{\bf s}}q({\bf s})\ln{\frac{[S_{\rm{tot}}]}{[M_{tot}]^{|\bf s|}}} \right).
\end{align*}
This expression can be re-written as
\begin{equation}
 \Delta \dot{G}=J_{\rm{tot}}\Delta G_{\rm{chem}} + J_{\rm{tot}}\Delta G_{\rm{inf}},
 \label{INFCHEM}
\end{equation}\\
with
\begin{align}
\Delta G_{\rm chem} &=  (|{\bf s}|-1) \Delta G^\plimsoll_{\rm dim} + \ln{[S_{\rm tot}]/[M_{\rm tot}]^{|{\bf s}|}}, \nonumber \\
\Delta G_{\rm{inf}} &= \sum_{{\bf s}}q({\bf s})\ln{\frac{p({\bf s})}{t(\bf s)}},
\label{dG_inf}
\end{align}
assuming for simplicity that all oligomers are of the same length. The first term in Eq.~\ref{INFCHEM} is the average chemical free-energy change of oligomerisation ignoring sequence, multiplied by the net rate of oligomer production. The second term is
information-theoretic: for non-negative net production of all oligomers \(J({\bf s}) = q({\bf s})J_{\rm tot} \geq 0\), \( q({\bf s})\) is the probability of picking a sequence \({\bf s}\) from the pool of net products and \(\Delta G_{\rm inf} = D(q||t) -D(q||p)\), where \(D(q||p) = \sum_{\bf s}q({\bf s})\log{\frac{q({\bf s})}{p({\bf s})}}\) is the  Kullback-Leibler divergence between \(q({\bf s})\) and \(p({\bf s})\). \(\Delta G_{\rm inf}\) reflects the sequence statistics of monomer and oligomer baths, and the sequence-dependence of net oligomer production. This splitting into chemical and informational terms holds for arbitrary oligomer lengths and sequence alphabets, and is the first result of this chapter.

\subsection{Is there necessarily a thermodynamic cost to accuracy?}

We return to our model, shown in Fig.~\ref{Fig1}, with two varieties of monomer in the template and copy. 

We assume constant bath concentrations, and calculate the flux \(J({\bf s})\) in steady state by analysing the Markov process corresponding to the states of a single template \cite{[{See supplemental material at}][{for additional data, derivations and discussion.}]SI}.

\begin{figure}[]
    \centering
    \includegraphics[scale=0.35]{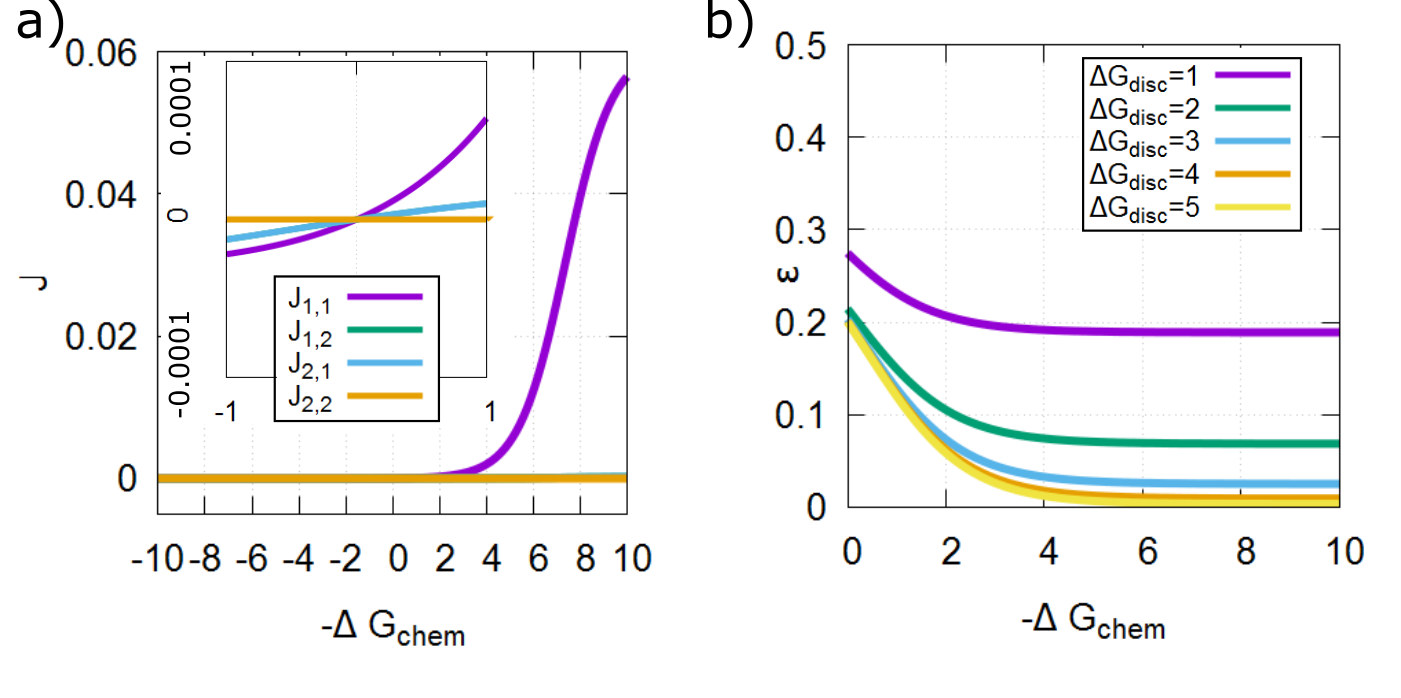}
        \caption{{\it A dimer copying model shows finite accuracy in the equilibrium limit}. (a)  The net production rate \(J({\rm s})\) against chemical driving \(-\Delta G_{\rm chem}\) for each of the four dimers 1,1, 1,2, 2,1 and 2,2 relative to a template of sequence 1,1, with \(\Delta G_{\rm disc}=\Delta G^\plimsoll_{w}- \Delta G^\plimsoll_{r}=10, \Delta G^\plimsoll_r+ \Delta G^\plimsoll_w = 0\) and unbiased monomer and oligomer baths. All \(J({\rm s})\) pass through zero at the equilibrium point \(\Delta G_{\rm chem}=0\) but quickly separate when \(\Delta G_{\rm chem}\neq0\) (inset). \(J({1,2})\) and \(J({2,1})\) overlap. At larger driving, accurate copies are preferentially produced. (b) Error fraction at which incorrect type 2 monomers are incorporated into dimers, \(\epsilon\), against \(\Delta G_{\rm chem}\) for various \(\Delta G_{\rm disc}\) and \(\Delta G_r+ \Delta G_w = 0\).  The system has finite accuracy, $\epsilon \neq 0.5$, as \(\Delta G_{\rm chem} \rightarrow 0^-\), with the accuracy dependent on \(\Delta G_{\rm disc}\).}
    \label{Fig2}
\end{figure}

In all previous work on infinite length templates there is an entropic cost to creating an accurate, low entropy copy. In templated self-assembly systems\cite{Bennett,Cady,Andrieux,Sartori1,Sartori2,esposito2010,EHRENBERG1980333,Johansson}, low entropy sequences can be compensated by favourable bonds between copy and template. In our previous work on templated copying which incorporates separation\cite{Ouldridge,Poulton} this entropic cost has to be paid for directly by the system. The general result in Eq.~\ref{INFCHEM}, however, suggests the accuracy of the created oligomer distribution doesn't directly affect the free energy required to create the distribution. 

To illustrate, let us first consider a template of sequence 1,1 coupled to  baths 
where all oligomers have the same  concentration \([S_{\rm tot}]/4\), and all monomers have the same concentration \([M_{\rm tot}]/2\). Without loss of generality we may choose our standard concentration so that \([S_{\rm tot}]/[M_{\rm tot}]^2=1\), and thus \(\Delta G_{\rm{chem}}=\Delta G^\plimsoll_{\rm{dim}}\).

In Fig.~\ref{Fig2}(a) we plot the net production rate of each dimer as a function of \(\Delta G_{\rm chem}\) at set \(\Delta G_{\rm disc}=\Delta G^\plimsoll_{r}-\Delta G^\plimsoll_{w}\). Equilibrium is at \(\Delta G_{\rm chem}=0\), with net creation for all dimers if \(\Delta G_{\rm chem}<0\)  and net destruction if \(\Delta G_{\rm chem}>0\). In Fig. 2(b) we plot \(\epsilon = \left(J(2,2)+\frac{1}{2}(J(1,2)+J(2,1))\right)/J_{\rm tot}\), the proportional rate at which incorrect monomers are incorporated into dimers. It is noticeable that while at exactly \(\Delta G_{\rm{chem}}=0\), \(\epsilon\) is undefined, as \(\Delta G_{\rm{chem}}\rightarrow 0^-\) we obtain \(\epsilon < 0.5\), implying non-zero accuracy.

\begin{figure}[h!]
    \centering
    \includegraphics[scale=0.35]{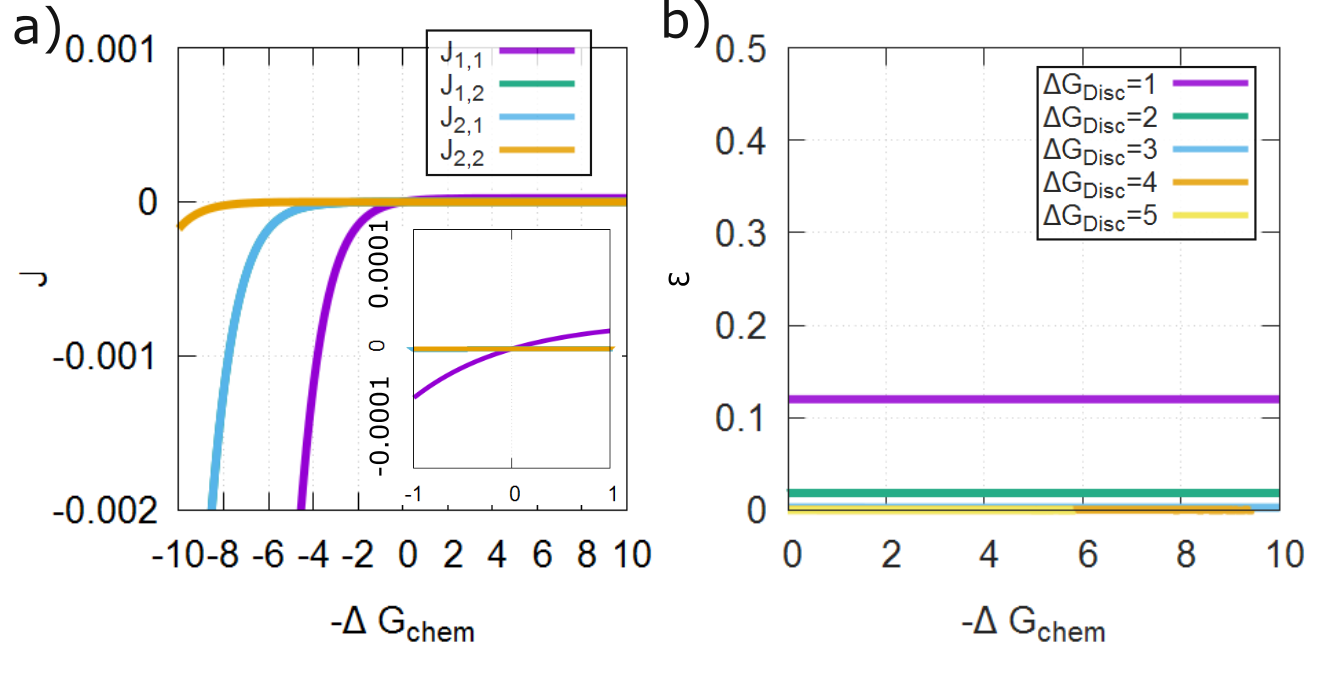}
        \caption{{\it The dimer copying model can show perfect accuracy in the equilibrium limit}. (a)  We plot the net production rate \(J_{\rm s}\) against chemical driving \(-\Delta G_{\rm chem}\) for each of the four dimers 1,1, 1,2, 2,1 and 2,2 relative to a template of sequence 1,1 and with \(\Delta G_{\rm disc}=\Delta G_{w}^{\plimsoll}- \Delta G_{r}^{\plimsoll}=2\), but \(\frac{\Delta G_{r}^{\plimsoll} + \Delta G_{w}^{\plimsoll}}{2}=5\). All oligomers have the same (fixed) concentration \([S_{\rm tot}]/4\), and all monomers have the same (fixed) concentration \([M_{\rm tot}]/2\). All \(J_{\rm s}\) pass through zero at the equilibrium point \(\Delta G_{\rm chem}=0\) but separate quickly (inset) at non-zero driving. Here accurate copies are preferentially produced but all fluxes are very low. (b) Error fraction at which incorrect type 2 monomers are incorporated into dimers, \(\epsilon\), against \(-\Delta G_{\rm chem}\) for various \(\Delta G_{\rm disc}\) with \(\frac{\Delta G_{r}^{\plimsoll} + \Delta G_{w}^{\plimsoll}}{2}=5\).  The system tends towards perfect accuracy, $\epsilon \rightarrow 0$, as \(\Delta G_{\rm disc} \rightarrow \infty\), even as the system tends to the equilibrium point \(\Delta G_{\rm chem}\rightarrow0^{-}\).}
    \label{SI1}
\end{figure}

Fig.~\ref{SI1} shows that when the average template binding free-energy of right and wrong monomers is increased, at fixed \(\Delta G_{\rm disc}=\Delta G_w - \Delta G_r\), the error remains low as the system tends towards equilibrium (\(\Delta G_{\rm chem}\rightarrow0^-\) for unbiased \(p({\bf s})\), \(t({\bf s})\) and \([S_{\rm tot}]/[M_{\rm tot}] =1\)). Indeed, \(\epsilon \rightarrow 0\) (perfect accuracy) as \(\Delta G_{\rm disc} \rightarrow \infty\) and \(\Delta G_{\rm chem} \rightarrow 0^{-}\). The error at \(\Delta G_{\rm chem}=0\) is still undefined, but the fluxes separate quickly after this point (inset Fig. \ref{SI1}a) to keep the error low. Due to the unstable bonds between copy and template, the system has a low flux for $\Delta G_{\rm chem}<0$.

The  minimal value of \(-\Delta G_{\rm{chem}}\) required for growth gives non-zero accuracy, and reversible processes can create a low entropy dimer sequence distribution at finite discrimination free-energy \(\Delta G_{\rm disc}\).
However, interactions between template and product do not persist; \(\Delta G_{\rm disc}\) does not feature in the overall thermodynamics, and cannot compensate for a low entropy state in equilibrium, as in templated self-assembly~\cite{Bennett,Cady,Andrieux,Sartori1,Sartori2,esposito2010,EHRENBERG1980333,Johansson}.
So is there no thermodynamic cost to accuracy in this setting? Consider \(\Delta G_{\rm inf}\) in Eq.~\ref{dG_inf}. In this case, \(p({\bf s})\) and \(t({\bf s})\) are unbiased and equal, and thus \(\Delta G_{\rm inf}=0\) for any \(q({\bf s})\) - even if only accurate copies are produced. Whenever the surrounding oligomers and monomers have no (or the same) sequence bias there is no extra thermodynamic cost to producing sequences of arbitrary accuracy. We emphasise the surprising result that you can get arbitrary accuracy at no extra cost {\it even if} the monomer distribution is unrelated to the template sequence.

\(\Delta G_{\rm inf}\) is generally non-zero, however, for systems with \(p({\bf s})\neq t({\bf s})\), {\it i.e.,} where the monomer and oligomer baths have different distributions. \(\Delta G_{\rm inf}\) is positive if the system produces sequences that are common in the oligomer bath \(p({\bf s})\) and rare in the monomer bath \(t({\bf s})\). The alternative representation \(\Delta G_{\rm inf} = D(q||t) - D(q||p)\) makes this fact particularly clear: the probability distribution of the creation fluxes \(q({\bf s})\) being similar to the monomer \(t({\bf s})\) bath makes the first term less positive and \(q({\bf s})\) being unlike the oligomer bath \(p({\bf s})\) makes the second term more negative.  Accuracy is therefore not directly constrained by thermodynamics in a general description of the full process of oligomer copying. Instead, there is a thermodynamic cost  to producing sequence distributions \(q({\bf s})\) that are closer to the oligomer sequences in the environment than a distribution of sequences obtained by randomly sampling monomers from the environment. 
This argument is the second main result of this chapter.


\section{Template copying as an inherently non-equilibrium information engine}
Unlike templated self-assembly systems~\cite{Bennett,Cady,Andrieux,Sartori1,Sartori2,esposito2010,EHRENBERG1980333,Johansson} in this system the template can act cyclically. It can therefore be thought of as an engine in the conventional thermodynamic sense, like a Carnot cycle. Here, the engine operates between the monomer and oligomer reservoirs. Its thermodynamics is set by the relationship between its input from and its output to these reservoirs relative to the chemical potential of those reservoirs. This is the physics encapsulated by our main result. 

\begin{figure}[]
    \centering
    \includegraphics[scale=0.35]{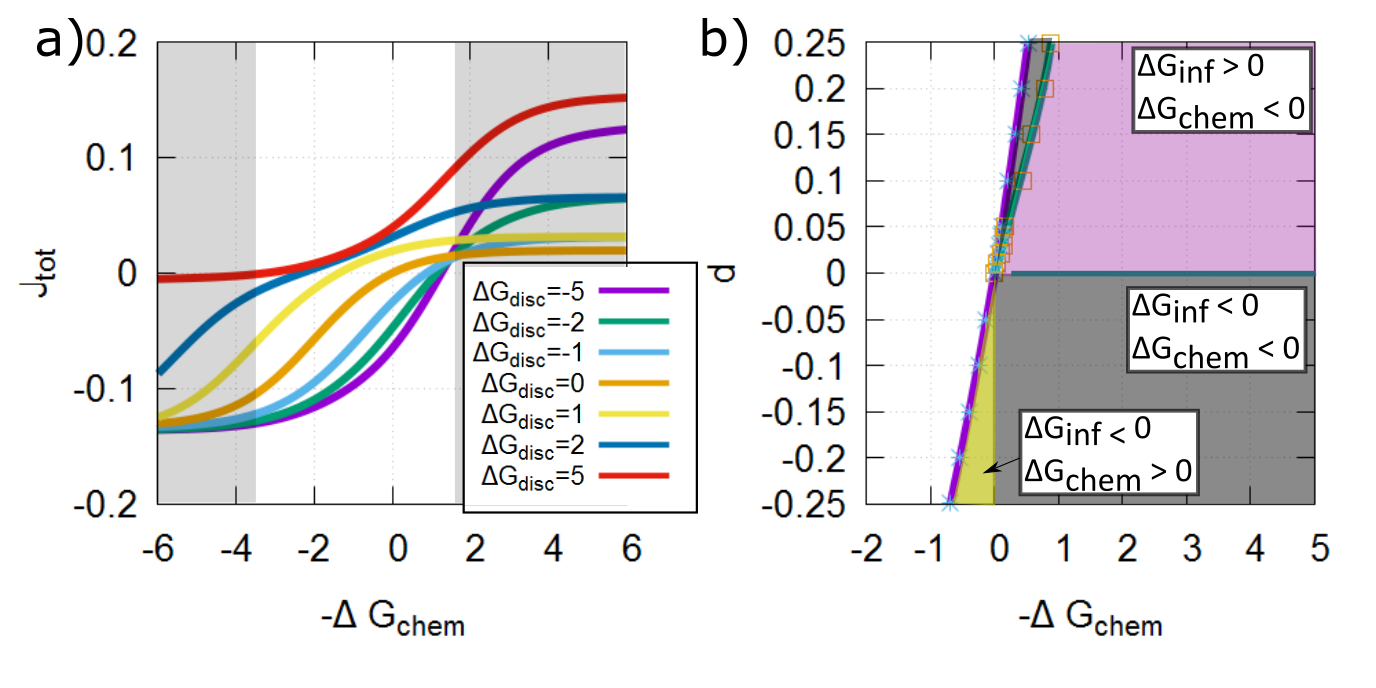}
    \caption{{\it Oligomer copying as an information engine with no equilibrium point}. (a) Total flux \(J_{\rm{tot}}\) against driving \(-\Delta G_{\rm chem}\) for a range of discrimination free energies \(\Delta G_{\rm disc}=\Delta G^\plimsoll_{w}-\Delta G^\plimsoll_{r}\), with
   \([1]=[2]=0.1\), \([1,1]=[1,2]=[2,1]=0.001\) and \([2,2]=0.1\).
    \(\Delta G_{\rm disc}\) is varied with \(\Delta G^\plimsoll_{r}+\Delta G^\plimsoll_{w}=0\) fixed. The point \(J_{\rm tot}=0\) at which there is no net dimerisation varies within the allowed white range despite the fact that  the overall dimerisation free-energy is independent of \(-\Delta G_{\rm disc}\). Specificity for  \({\bf s}_{\rm min} = 1,1\) 
    makes growth easier and pushes $J_{\rm tot}=0$ to the lower limit, and specificity for \({\bf s}_{\rm max}=2,2\) has the opposite effect. (b) Phase plot of the information engine. Here the leftmost purple boundary is the transition from \(J_{\rm tot}\) negative to positive. Here we fix \(\Delta G_{\rm disc}=5\), \([S_{\rm tot}]=1\), \([M_{\rm tot}]=1\), \(t(s)=0.25\) for all \(s\) and vary \(\Delta G_{\rm chem}\). We further vary \(p(s)=0.25+d,0.25,0.25,0.25-d\) by varying \(d\). There is a regime in which chemical work is used to specifically produce sequences of high free-energy and a regime in which specific production of low free-energy sequences is used to drive oligomerisation against a chemical load.}
    \label{Fig3}
\end{figure}

Unusually, this system can couple to so many reservoirs that it is generally impossible to find an equilibrium point. 
To illustrate, consider a set of systems with \(p({\bf s})\neq t({\bf s})\), {\it i.e.}, where the monomer and polymer baths are different. Here, there is no equilibrium point at which all fluxes are zero because the baths are out of equilibrium with each other. There is instead a range of \(\Delta G_{\rm chem}\)  over which \(J_{\rm tot} =0\) could occur, depending on which sequences best couple to the template. The most positive possible \(\Delta G_{\rm chem}\) at which \(J_{\rm tot} =0\) occurs is observed when a system  specifically produces the sequence \({\bf s}_{\rm min}\), where \({\bf s_{\rm min}}\) minimises \(t({\bf s})/p({\bf s})\). The most negative is when the system specifically produces the sequence \({\bf s}_{\rm max}\), where \({\bf s_{\rm max}}\) maximises \(t({\bf s})/p({\bf s})\). \({\bf s}_{\rm min}\) is intuitively the sequence most like the monomer baths and least like the polymer baths and \({\bf s}_{\rm max}\) is the opposite. In Fig.~\ref{Fig3}(a), we vary \(\Delta G_{\rm disc}\) for a system heavily thermodynamically biased towards creating accurate copies by the baths. When \(\Delta G_{\rm disc}>0\), and the system is also kinetically biased towards creating accurate copies and \(J_{\rm tot}=0\) for a more positive \(\Delta G_{\rm chem}\) than if \(\Delta G_{\rm disc}<0\). Copying accurately can thus either make production of polymers thermodynamically easier or harder, depending on the environment. This fact is true even for an unbiased pool of monomers. 


We define \(q_{\rm min}({\bf s})\) as the \(q({\bf s})\) that results in the most negative value of \(\Delta G_{\rm inf}\), and \(q_{\rm max}({\bf s})\) which maximises \(\Delta G_{\rm inf}\). When the probability distribution of fluxes \(q({\bf s})\) is close to \(q_{\rm min}({\bf s})\) it is possible for a negative \(\Delta G_{\rm inf}\) to overcome a positive \(\Delta G_{\rm chem}\). Equally, a more negative \(\Delta G_{\rm chem}\) allows for a \(q({\bf s}) \approx q_{\rm max}({\bf s})\) with positive \(\Delta G_{\rm inf}\). From this perspective, thinking about the monomer and oligomer reservoirs as a single collective environment, we can describe the template as an engine that trades chemical and information-based free energy in the environment against each other \cite{mcgrath2017biochemical,PhysRevLett.111.010602}. 

The second law implies that the rate of change of free energy \(\Delta \dot{G}\) is negative. Thus, from Eq.~\ref{INFCHEM}, there are three possible regimes for this informational engine, illustrated in Fig. \ref{Fig3}(b). If \(\Delta G_{\rm chem}<0\) and \(\Delta G_{\rm inf}>0\) then the system channels chemical work through a specific copying mechanism to store free energy in a distribution of outputs closer to the polymer bath \(p({\bf s})\) than the monomer bath \(t({\bf s})\), with an efficiency \(\eta=\frac{\Delta G_{\rm inf}}{-\Delta G_{\rm chem}} \leq 1\). In our case, \(\eta\) reaches a maximum of \(\sim 0.3\) when \(p({\bf s})\) is heavily biased towards accurate copies of the template and \(\Delta G_{\rm chem}\) is small and negative. In the case where \(\Delta G_{\rm chem}>0\) and \(\Delta G_{\rm inf} <0\), the system generates outputs closer to the monomer baths \(t({\bf s})\) than the polymer baths \(p({\bf s})\), expending information to compensate for an unfavourable chemical work term. Here the efficiency \(\eta=\frac{\Delta G_{\rm chem}}{-\Delta G_{ \rm inf}} \leq 1\) reaches a maximum of \(~0.15\) when \(p({\bf s})\) is heavily biased against accurate copies of the template and \(\Delta G_{\rm chem}\) is small and positive. The final case, in which both \(\Delta G_{\rm chem}\leq 0\) and \(\Delta G_{\rm inf} \leq 0\), is less interesting as the system both spends chemical free energy and generates outputs close to \(q({\bf s})=q_{min}({\bf s})\).

%


%

\section{Kinetic convergence on thermodynamic constraints for infinite-length polymers}

\begin{figure}
    \centering
    \includegraphics[scale=0.35]{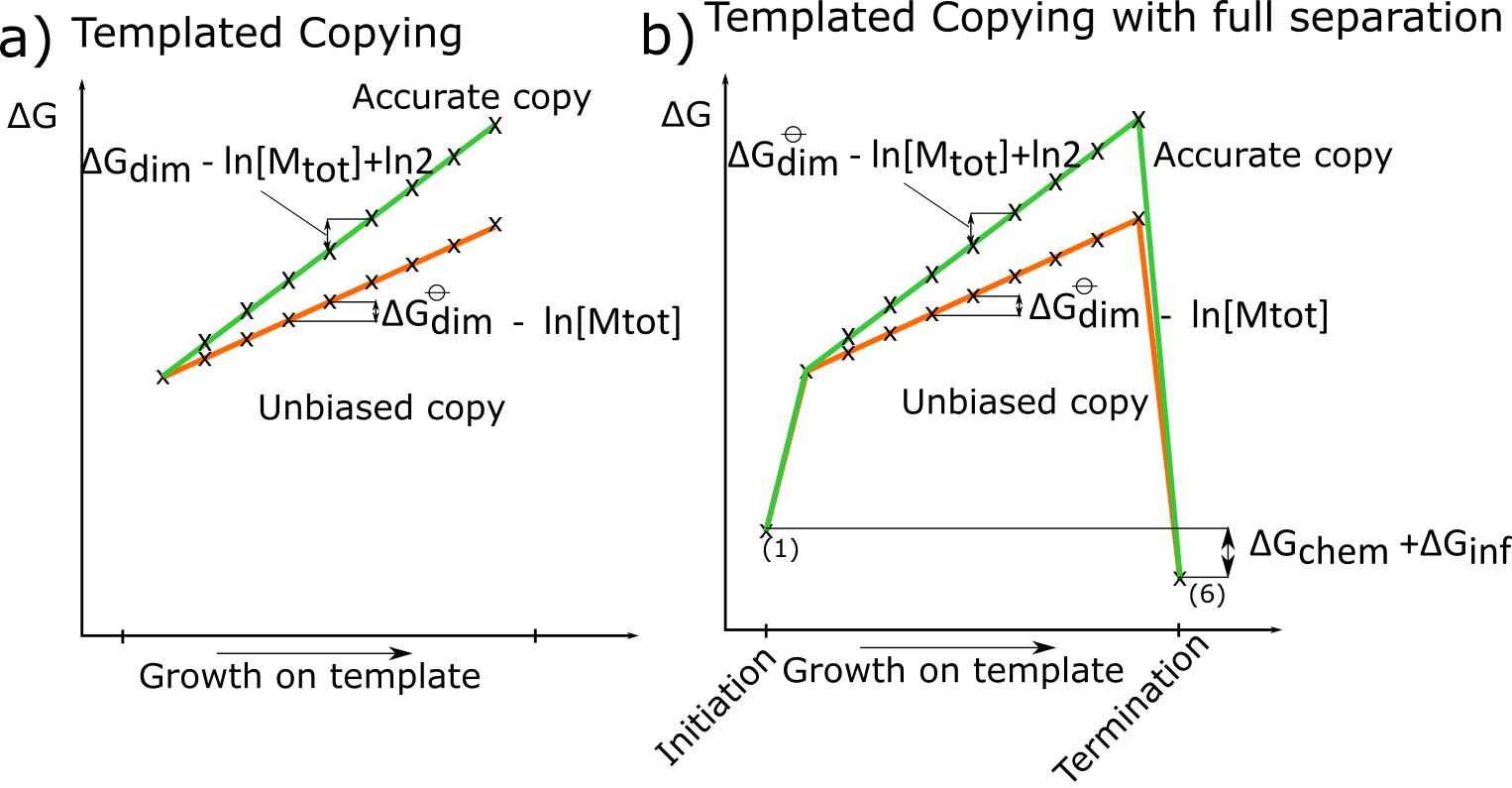}
    \caption{Free-energy profiles for (a) a templated copying process on an infinite lengths template and (b) a finite length template for the process of going from $G_{0,N}$ to $G_{0,N+1}$.. The gradient of the free-energy profile during the bulk of the copying process is set by the free-energy of dimerisation, the chemical driving due to the relative concentrations of the monomer and oligomer baths and the accuracy of the oligomer; an accurate oligomer required an extra \(\ln{2}\) to add a  monomer in the bulk more than a unbiased oligomer. However in the case of a finite length template, the initiation and termination steps can provide a theoretically unlimited adjustment to the overall free-energy of the process. In some cases creating the most accurate oligomer can be thermodynamically favourable compared to other oligomers.}
    \label{Lastfig}
\end{figure}

Our results apply to polymers of arbitrary finite length. So are the thermodynamic constraints derived when considering continuous separation of copy and template during polymerisation, but neglecting initiation and termination, invalid? For example, what becomes of the claim that inputs of higher chemical free energy (more chemical work) are required for an accurate copy than an inaccurate one in templated copying \cite{Poulton}? 

To answer this question, we return to the free-energy profiles of Fig.~\ref{NewFig1}, reproducing the plot for templated copying in Fig.~\ref{Lastfig}\,(a). In Fig.~\ref{Lastfig}\,(b), we illustrate schematically the effect of considering initiation and termination via a model such as Fig.~\ref{Diagrams}\,(c). Firstly, it is no longer sufficient to label the states purely by the length of copy attached to the template; a state with one more oligomer created and released to the baths has a distinct free energy. We therefore consider $G_{l,N}$, the free energy of the copy, template and surrounding baths restricted to states with a copy of length $l$ and in which $N$ oligomers have been produced.  Secondly, since the template is not infinitely long, the overall thermodynamic favourability of the process is determined by $G_{0,N+1}- G_{0,N}$, rather than the slope of the free-energy profile while template-attached.

Initiation and termination provide a theoretically unlimited adjustment to the overall \(\Delta G\), since the entropy of the copy itself is no longer relevant once it has detached from the template and mixed with the environment, but the relationship of its sequence to the oligomer bath disribution $p({\bf s})$ is. An arbitrarily unfavourable polymerisation process on the template can be made favourable with the right concentration of products, despite an uphill free-energy profile while attached to the template. 

Importantly, however, if template-attached growth is unfavourable ({\it i.e.}, the free-energy profile is uphill during the copy process), it will be kinetically suppressed by a large free-energy barrier even if oligomer production is favourable overall. Here, barrier height grows proportionally to oligomer length, suggesting that the {\it thermodynamic} constraints derived for infinite-length polymers should become increasingly prohibitive {\it kinetic} effects for sufficiently long oligomers.
To probe this hypothesis we consider a kinetic model for the growth and destruction of oligomers of arbitrary fixed length, by extending the model of Ref.~\cite{Poulton} so that it includes initiation and termination (Fig.~\ref{Diagrams}\,(c)). 

\subsection{Model of copy production for oligomers of length \(|s|>2\)}

The model of Ref.~\cite{Poulton} considers a single template consisting of a series of monomers \({\bf n}=n_{1},...,n_{L}\),  with $L \rightarrow \infty$. Growing on the template is a single copy oligomer \({\bf s}=s_{1},...,s_{l}\), (\(l\leq L\)). Inspired by transcription and translation, the model describes a copy that detaches from the template as it grows. Transitions in the model are whole steps in which a single monomer is added or removed from the copy's leading edge, potentially encompassing many individual chemical sub-steps (for ease of discussion we assume that the overall dynamics is well modelled by a single instantaneous step, but that is not essential for our conclusions). As illustrated in Figure \ref{Diagrams}(c), after each step there is only a single inter-polymer bond at position \(l\), between $s_l$ and $n_l$. As a new monomer joins the copy at position \(l+1\), the bond position \(l\) is broken.

We augment this existing model by restricting the template to a length $L$, and considering the possibility of an empty template as shown in Fig. \ref{Diagrams}c. Monomers can bind to or detach from the first site of an otherwise empty template, and full-length oligomers can bind to and detach from the final site of a template. We assume that at most one copy is bound to the template at any time.

As in the dimerisation model considered earlier in the text, we shall consider a template polymer \({\bf n}\) made entirely of monomers of type 1. Given the assumed symmetry between the interactions of the two monomer types, and equal concentrations of the monomer baths as used throughout this work, the results apply equally well to any single template sequence. Monomers of type 1 in the copy can simply be interpreted as correct matches and monomers of type 2 as incorrect matches for any template sequence \({\bf n}\). 

Having defined the model's state space, we now consider state free energies. By analogy with the dimerisation model defined in section \ref{sec:model}, we define \(\Delta G^\plimsoll_{\rm{dim}}\) as the free-energy change of adding any specific monomer to the end of the copy oligomer at standard concentrations, ignoring interactions with any template. 
The environment contains baths of monomers; a monomer of type \(s\) has a constant concentration \([s]\) relative to the standard concentration. 
The chemical free-energy change for the transition between any specific sequence \(s_{1},...,s_{l}\) and any specific sequence \(s_{1},...,s_{l+1}\), ignoring any contribution from interactions with the template, is then \(\Delta G^\plimsoll_{\rm{dim}} - \ln [s_{l+1}]\).
We then consider the effect of specific interactions with template. Analogously to before, we define $\Delta G^\plimsoll_{r/w}$ as the standard free energy of binding for matched/mismatched monomers and the template at standard concentrations, with $ \Delta G_{\rm disc}=\Delta G^\plimsoll_{w}-G^\plimsoll_{r}$ quantifying the additional stability of correct matches. Only the leading monomer is assumed to interact with the template. The overall chemical free-energy change of a given transition is then easily calculable. Each copy extension step from $l$ to $l+1$ is associated with a chemical free-energy change of \(\Delta G^\plimsoll_{\rm{dim}} - \ln [s_{l+1}] +  \Delta G_{\rm disc}\) (leading copy monomer goes from $r \rightarrow w$); \(\Delta G^\plimsoll_{\rm{dim}} - \ln [s_{l+1}] +  0\) (for $w \rightarrow w$ or $r \rightarrow r$); or \(\Delta G^\plimsoll_{\rm{dim}} - \ln [s_{l+1}] -  \Delta G_{\rm disc}\) (for $w \rightarrow r$). These free-energy changes are so simple because each copy extension step doesn't increase the number of copy monomers interacting with the template (Fig.~\ref{Diagrams}\,(c)). Instead, the result is that a different single copy monomer is bound to the template.  

A monomer $s_1$ binding to the first site of the template from solution is associated with a free energy change of \(G^\plimsoll_{r/w} - \ln[s_1]\), depending on whether it is a correct match to the first site of the template. Similarly, an oligomer ${\bf s}$ binding from solution is associated with a free energy change of \(G^\plimsoll_{r/w} - \ln[{\bf s}]\), depending on whether it is a correct match to the final site of the template.



These free-energy changes constrain the relative propensities of forwards and backwards transitions\cite{ouldridge2018importance}, but a range of kinetic models satify these constraints. In the temporary thermodynamic discrimination model of Ref.~\cite{Poulton}, polymerization steps involving the addition of monomer $s_i$ are assumed to occur with the same rate $k[s_i]$, and sequence-based discrimination occurs in the backwards step. We use that parameterization here; in addition, we assume the binding of individual monomers \(s_i\) or oligomers \({\bf s}\) to the an empty template also has a rate of $k[s_i]$ or $k[{\bf s}]$, with the off rates fixed by the free-energy change of transition. These assumptions fix all rates, which are given in \cite{[{See supplemental material at}][{for additional data, derivations and discussion.}]SI}.

%

\begin{figure}[]
\includegraphics[scale=0.25]{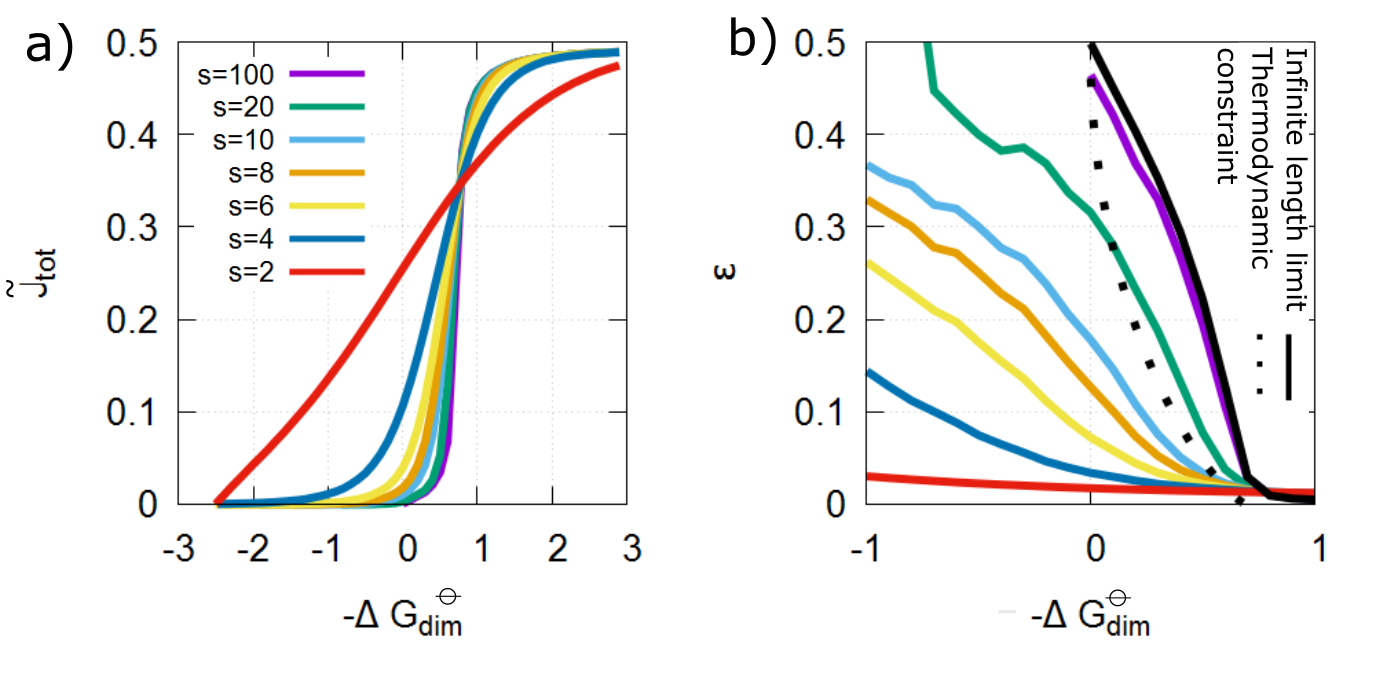}
\caption{{\it Thermodynamic restrictions in an infinite-length model become kinetic restrictions for longer oligomers.} (a) Net flux per unit empty template \(\tilde{J}_{tot}\) of oligomer production and (b) the net fraction of error creation  for a range of lengths \(|{\bf s}|\). We vary \(\Delta G_{\rm dim}^\plimsoll\), with \([M_{\rm tot}]=1\), \([S_{\rm tot}]\) chosen to give \(\Delta G_{\rm chem} =0\) at \(\Delta G_{\rm dim}^{\plimsoll} = -5\), \(p(s)=t(s)\) unbiased and \(\Delta G_{\rm disc}=8\) at \(\Delta G_r+\Delta G_w =0\). Also plotted in (d) is the  thermodynamic constraint on accuracy for an infinite length polymer, set by requiring a non-positive slope of the free-energy profile, \(-\epsilon\ln{\epsilon}-(1-\epsilon)\ln{(1-\epsilon)}\geq  \Delta G_{\rm dim}^\plimsoll+\ln2\),and the actual error rate obtained for an infinite-length copy for these parameters \cite{Poulton}. Short oligomers  overcome kinetic barriers to produce copies with \(\Delta G_{\rm dim}^\plimsoll>0\) and \(\epsilon\) below the infinite-length limit and thermodynamic constraint.} 
\label{fig4}
\end{figure}


To analyse the model we perform stochastic simulations for a range of lengths  \(|{\bf s}|\), varying the free-energy released by backbone formation \(\Delta G_{\rm dim}^\plimsoll\) while keeping all other parameters fixed. We use unbiased \(m({s})\) and \(p({\bf s})\), set $[M_{\rm tot}] =1$ and choose $[S_{\rm tot}]$ so that \(\Delta G_{\rm chem}=0\) at \(\Delta G_{\rm dim}^{\plimsoll}=5\); growth is thermodynamically favourable for all sequences when \(\Delta G_{\rm dim}^{\plimsoll}<5\). However, the slope of the on-template free-energy profile of an unbiased sequence, \(\Delta G_{\rm dim}^\plimsoll -\ln [M_{\rm tot}]\),  is positive for \(\Delta G_{\rm dim}^{\plimsoll}>0\). For \(5> \Delta G_{\rm dim}^{\plimsoll} > 0\), therefore, on-template polymerisation is thus a kinetic barrier to formation of a thermodynamically favourable product. 

We calculate the total flux per empty template \(\tilde{J}_{\rm tot}=\frac{J_{\rm tot}}{P_{\rm empty}}\) in the steady state and plot in Fig. \ref{fig4}. For short oligomers, non-negligible \(\tilde{J}_{\rm tot}\) is observed in the region \(5> \Delta G_{\rm dim}^{\plimsoll} > 0\), indicative of the system surmounting the free energy barriers associated with unfavourable growth on the template. However, as oligomers get longer, the barriers get larger and the kinetics is slowed. Both forward and backwards contributions to \(\tilde{J}_{\rm tot}\) are vanishingly small unless \(\Delta G_{\rm dim}^{\plimsoll} < 0\) for \(|s|=100\). 


Kinetic barriers not only control overall the production flux per empty template \(\tilde{J}_{\rm tot}\), but also error incorporation. The on-template production of an accurate copy  has a more positive slope in its free-energy profile than an unbiased sequence (Fig.~\ref{Lastfig}(b)). For an infinite-length polymer, this fact provides a thermodynamic constraint on accuracy for \(0>\Delta G_{\rm dim}^{\plimsoll}>-\ln 2\) \cite{Poulton}. We plot the fraction of net incorporated monomers that do not match the template, \(\epsilon\), in Fig.~\ref{fig4}(b), alongside the thermodynamic constraint on $\epsilon$ for infinite-length polymers and the actual error rate obtained for this specific model in the infinite-length limit \cite{Poulton}. Short oligomers can overcome kinetic barriers and beat both the thermodynamic bound and the accuracy obtained in the infinite-length limit; longer oligomers approach the limiting behaviour slowly, with significant differences even at length 20.

\section{Conclusion}

In this article we have investigated the copying of finite-length oligomers, with explicit focus on initiation and termination. Copying creates correlations between copy and template sequences, but the mixing of the products with oligomers in the environment means that the information between copy and template sequences is not thermodynamically exploitable \cite{Ouldridge}.

Accurate copying creates a low entropy sequence. In templated self-assembly, this low entropy is costly, but can be compensated for and even favoured in equilibrium due to stabilising interactions of specific copy/template bonds. If these bonds are transient, however, as in the production of a true copy, this compensation is impossible in the overall thermodynamics of the process. In an infinite-length model of continuous detachment from behind the leading edge of the copy \cite{Poulton}, the low entropy of the product implies a non-equilibrium state. A reduction in the entropy of the copy of $\sim \ln 2$ per monomer relative to a random sequence must be formed without a compensating discrimination free energy that scales with the length of the copy.

The argument that transient bonds prevent the template from biasing the system towards a low entropy state in equilibrium \cite{Poulton} remains valid once full initiation and termination are considered. However, once the copy has fully detached and mixed with an environment, including oligomers of other sequences, the low entropy of the specific copy oligomer is thermodynamically irrelevant. Since that individual copy can no longer be identified without prior measurement of its sequence, what matters thermodynamically is the contribution of that copy to the entropy of the oligomer distribution as a whole, not the sequence entropy of the sequence actually produced. This change in the thermodynamic significance of the copy sequence itself is manifest as the arbirtarily-large offset in the final step of the free energy profile, shown in Fig. \ref{Lastfig}.

Thus, as we have shown, the overall thermodynamics of the full copy process does not explicitly depend on accuracy. Instead, the surrounding concentrations of oligomers and monomers set the thermodynamic constraints. Creating outputs that resemble the surrounding oligomers is costly, as is creating outputs unlike the input monomer baths. Arbitrary accuracy can be free-energetically neutral or even actively favourable if the oligomer baths are biased towards other sequences.

However, accuracy does play a role indirectly. Firstly, mixing with other oligomers is the final step, and therefore its thermodynamic consequences are irrelevant whilst a copy is growing on the template. Whilst attached to the template -- even if only by the leading site -- the copy is distinct from the surround oligomer pool and its low entropy is exploitable. For an infinite-length polymer, the associated costs \cite{Poulton} set absolute limits on what is possible. For finite length oligomers, they instead manifest as kinetic barriers; longer oligomers  have larger barriers and thus their kinetics converges on the behaviour dictated by the thermodynamic constraints.

Secondly, templates will typically influence their environment. If a template sets its own oligomer environment, \(p({\bf s}) = q({\bf s})\), \(\Delta G_{\rm inf}= D(q||t) \), which reduces to the entropy difference between \(t(\bf{s})\) and  \(q({\bf s})\) if \(t(\bf{s})\) is unbiased. In this case no information is lost upon mixing and accurate copying incurs a cost; the limits derived in \cite{Poulton} hold exactly. In general, there is no reason to suppose that \(p({\bf s}) = q({\bf s})\). As in a cell, other templates and differential degradation rates may be relevant in setting \(p({\bf s}) \). Nonetheless, particularly for longer oligomers, sequences common in q({\bf s}) are likely to be over-represented in \(p({\bf s}) \). 

 If many identical templates are present, then the environmental \(p({\bf s})\) will likely be more strongly peaked, and the cost of accuracy higher, than in a system with many distinct templates. Moreover, any template in an environment dominated by the copies of another will experience a relative thermodynamic advantage. This effect would act as a form of ``rubber banding" in evolutionary competition among minimal replicators, and favour virus-like templates invading new environments.
 
 In this work we have derived general results, assuming that there are no kinetic proofreading cycles that would consume a variable amount of fuel for each polymer produced and contribute an extra term to the overall free-energy change (Eq.~\ref{INFCHEM}). The central conclusions in this manuscript would be largely unaffected, however. The existence of kinetic proofreading cycles would effectively provide an unpredictable adjustment to $\Delta G_{\rm chem}$ for each oligomer produced; the role of $\Delta G_{\rm inf}$ and the copy accuracy in determining overall thermodynamics would be largely unchanged.

\bibliography{aipsamp}

\clearpage


\widetext
\begin{center}
\textbf{\large Supplemental Materials: Edge-effects dominate the fundamental thermodynamics of molecular copying for finite-length oligomers}
\end{center}

\setcounter{equation}{0}
\setcounter{figure}{0}
\setcounter{table}{0}
\setcounter{page}{1}
\setcounter{section}{0}
\makeatletter
\renewcommand{\theequation}{S\arabic{equation}}
\renewcommand{\thefigure}{S\arabic{figure}}
\renewcommand{\bibnumfmt}[1]{[S#1]}

\section{Finding the fluxes through the network using a transition matrix}

Fig. \ref{Fig1} defines a Markov process for the states of a single template of type \(1,1\). Note that we assume the dimers have a directionality (like biopolymers such as nucleic acids and polypeptides), so that 1,2 is distinct from 2,1. Below, we use ``left" to refer to the first site and ``right" to the second, for consistency with Fig. \ref{Fig1}. 

The available states are as follows; 
\small
\begin{itemize}
    \item State 0: the empty template (shown in five different locations in Fig. \ref{Fig1}, the four outer edges and the centre).
    \item State 1: the template with an incorrect monomer (2) on its left side.
    \item State 2: the template with an incorrect monomer (2) on its right side.
    \item State 3: the template with an correct monomer (1) on its left side.
    \item State 4: the template with an correct monomer (1) on its right side.
    \item State 5: the template with an incorrect monomer (2) on its left side and a correct monomer (1) on its right side.
    \item State 6: the template with an incorrect monomer (2) on its left side and an incorrect monomer (2) on its right side.
    \item State 7: the template with a correct monomer (1) on its left side and an incorrect monomer (2) on its right side.
    \item State 8: the template with a correct monomer (1) on its left side and a correct monomer (1) on its right side.
    \item State 9: the template with a \(2,2\) dimer attached to it.
    \item State 10: the template with a \(1,2\) dimer attached to it.
    \item State 11: the template with a \(1,1\) dimer attached to it.
    \item State 12: the template with a \(2,1\) dimer attached to it.
\end{itemize}
\normalsize

Using these states we can set up a rate matrix $K$ where $K_{xy}$ gives the transitions out of state \(x\) and into state \(y\):
\small
\begin{frame}{}
\footnotesize
\setlength{\arraycolsep}{2.5pt}
\medmuskip = 1mu 
\[\left(
\begin{array}{ccccccccccccccccc}
-X_{0} & [2]k_{\rm on} & [2]k_{\rm on} & [1]k_{\rm on} & [1]k_{\rm on} & 0 & 0 & 0 & 0 & 2[2,2]k_{\rm on} & 2[1,2]k_{\rm on} & 2[1,1]k_{\rm on} & 2[2,1]k_{\rm on}\\
k_{\rm on}e^{\Delta G_{w}} & -X_{1} & 0 & 0 & 0 & [1]k_{\rm on} & [2]k_{\rm on} & 0 & 0 & 0 & 0 & 0 & 0\\
k_{\rm on}e^{\Delta G_{w}} & 0 & -X_{2} & 0 & 0 & 0 & [2]k_{\rm on} & [1]k_{\rm on} & 0 & 0 & 0 & 0 & 0\\
k_{\rm on}e^{\Delta G_{r}^{\plimsoll}} & 0 & 0 & -X_{3} & 0 & [2]k_{\rm on} & 0 & 0 & [1]k_{\rm on} & 0 & 0 & 0 & 0\\
k_{\rm on}e^{\Delta G_{r}^{\plimsoll}} & 0 & 0 & 0 & -X_{4} & 0 & 0 & [1]k_{\rm on} & [2]k_{\rm on} & 0 & 0 & 0 & 0\\
0 &k_{\rm on}e^{\Delta G_{r}^{\plimsoll}} & 0 & k_{\rm on}e^{\Delta G_{w}^{\plimsoll}} & 0 & -X_{5} & 0 & 0 & 0 & 0 & 0 & 0 & k_{\rm cat}\\
0& k_{\rm on}e^{\Delta G_{w}^{\plimsoll}} & k_{\rm on}e^{\Delta G_{w}^{\plimsoll}} & 0 & 0 & 0 & -X_{6} & 0 & 0 & k_{\rm cat} & 0 & 0 & 0\\
0 & 0 & k_{\rm on}e^{\Delta G_{r}^{\plimsoll}} & 0 & k_{\rm on}e^{\Delta G_{w}^{\plimsoll}} & 0 & 0 & -X_{7} & 0 & 0 & k_{\rm cat} & 0 & 0\\
0 & 0 & 0 & k_{\rm on}e^{\Delta G_{r}^{\plimsoll}} & k_{\rm on}e^{\Delta G_{r}^{\plimsoll}} & 0 & 0 & 0 & -X_{8} & 0 & 0 & k_{\rm cat} &  0\\
2k_{\rm on}e^{\ddagger} & 0 & 0 & 0 & 0 & 0 & k_{\rm cat} & 0 & 0 & -X_{9} & 0 & 0 & 0\\
2k_{\rm on}e^{*} & 0 & 0 & 0 & 0 & 0 & 0 & k_{ \rm cat} & 0 & 0 & -X_{10} & 0 & 0\\
2k_{\rm on}e^{\dagger} & 0 & 0 & 0 & 0 & 0 & 0 & 0 & k_{\rm cat} & 0 & 0 & -X_{11} & 0\\
2k_{\rm on}e^{*} & 0 & 0 & 0 & 0 & k_{\rm cat} & 0 & 0 & 0 & 0 & 0 & 0 & -X_{12}\\
\end{array}\right),\]
\end{frame}
\normalsize
where \(* = G^\plimsoll_{\rm dim} + \Delta G^\plimsoll_{r} + \Delta G^\plimsoll_{w}\), \(\dagger = G^\plimsoll_{\rm dim} + 2\Delta G^\plimsoll_{r}\) and \(\ddagger=\Delta G^\plimsoll_{\rm dim} + 2\Delta G_{w}^\plimsoll\). \(X_{x}\) is the sum over all the other terms in row \(x\). Now we can solve for the steady state \(\pi\) by finding the appropriate left-eigenvector \(\pi K = 0\). 

Next we consider the probability of the system creating either a \(1,1\), \(1,2\), \(2,1\) or \(2,2\) dimer, or destroying a dimer into component monomer parts, given an initial state with either a monomer or dimer bound to the template. We thus split the empty state (state 0) into five destination states, as follows: 13) the empty state having just released a monomer (corresponding to destruction), 14) the empty state having just released a \(1,1\) dimer, 15) the empty state having just released a \(1,2\) dimer, 16) the empty state having just released a \(2,1\) dimer and 17) the empty state having just released a \(2,2\) dimer. We put these states on the end of the list of states above, and delete the original empty state 0. Treating those states as distinct absorbing states, we can calculate the probability of reaching any one of them first, given a specific staring point\cite{peters2017reaction}. To do this we use the following  transition matrix that describes the discrete-time process embedded in the continuous-time model of dimerisation. This embedded discrete time process describes the sequence of states visited, without reference to the time taken. The matrix takes the form

\begin{frame}{}
\footnotesize
\setlength{\arraycolsep}{2.5pt}
\medmuskip = 1mu 
\[\left(
\begin{array}{c|c}
M & A\\
\cline{1-1}\cline{2-2}
0 & I
\end{array}\right),\]
\end{frame}
\normalsize

where \(M\) is
\begin{frame}{}
\footnotesize
\setlength{\arraycolsep}{2.5pt}
\medmuskip = 1mu 
\[\left(
\begin{array}{cccccccccccc}
0 & 0 & 0 & 0 & \frac{[1]k_{\rm on}}{N_{1}} & \frac{[2]k_{\rm on}}{N_{1}} & 0 & 0 & 0 & 0 & 0 & 0\\
0 & 0 & 0 & 0 & 0 & \frac{[2]k_{\rm on}}{N_{2}} & \frac{[1]k_{\rm on}}{N_{2}} & 0 & 0 & 0 & 0 & 0\\
0 & 0 & 0 & 0 & \frac{[2]k_{\rm on}}{N_{3}} & 0 & 0 & \frac{[1]k_{\rm on}}{N_{3}} & 0 & 0 & 0 & 0\\
0 & 0 & 0 & 0 & 0 & 0 & \frac{[1]k_{\rm on}}{N_{4}} & \frac{[2]k_{\rm on}}{N_{4}} & 0 & 0 & 0 & 0\\
\frac{k_{\rm on}e^{\Delta G_{r}^{\plimsoll}}}{N_{5}} & 0 & \frac{k_{\rm on}e^{\Delta G_{w}^{\plimsoll}}}{N_{5}} & 0 & 0 & 0 & 0 & 0 & 0 & 0 & 0 & \frac{k_{\rm cat}}{N_{5}}\\
\frac{k_{\rm on}e^{\Delta G_{w}^{\plimsoll}}}{N_{6}} & \frac{k_{\rm on}e^{\Delta G_{w}^{\plimsoll}}}{N_{6}} & 0 & 0 & 0 & 0 & 0 & 0 & \frac{k_{\rm cat}}{N_{6}} & 0 & 0 & 0\\
0 & \frac{k_{\rm on}e^{\Delta G_{r}^{\plimsoll}}}{N_{7}} & 0 & \frac{k_{\rm on}e^{\Delta G_{w}^{\plimsoll}}}{N_{7}} & 0 & 0 & 0 & 0 & 0 & \frac{k_{\rm cat}}{N_{7}} & 0 & 0\\
0 & 0 & \frac{k_{\rm on}e^{\Delta G_{r}^{\plimsoll}}}{N_{8}} & \frac{k_{\rm on}e^{\Delta G_{r}^{\plimsoll}}}{N_{8}} & 0 & 0 & 0 & 0 & 0 & 0 & \frac{k_{\rm cat}}{N_{8}} &  0\\
0 & 0 & 0 & 0 & 0 & \frac{k_{\rm cat}}{N_{9}} & 0 & 0 & 0 & 0 & 0 & 0\\
0 & 0 & 0 & 0 & 0 & 0 & \frac{k_{ \rm cat}}{N_{10}} & 0 & 0 & 0 & 0 & 0\\
0 & 0 & 0 & 0 & 0 & 0 & 0 & \frac{k_{\rm cat}}{N_{11}} & 0 & 0 & 0 & 0 \\
0 & 0 & 0 & 0 & \frac{k_{\rm cat}}{N_{12}} & 0 & 0 & 0 & 0 & 0 & 0 & 0
\end{array}\right),\]
\end{frame}
\normalsize
and \(T\) is
\begin{frame}{}
\footnotesize
\setlength{\arraycolsep}{2.5pt}
\medmuskip = 1mu 
\[\left(
\begin{array}{ccccc}
\frac{k_{\rm on}e^{\Delta G_{w}^{\plimsoll}}}{N_{1}} & 0 & 0 & 0 & 0 \\
\frac{k_{\rm on}e^{\Delta G_{w}^{\plimsoll}}}{N_{2}} & 0 & 0 & 0 & 0 \\
\frac{k_{\rm on}e^{\Delta G_{r}^{\plimsoll}}}{N_{3}} & 0 & 0 & 0 & 0 \\
\frac{k_{\rm on}e^{\Delta G_{r}^{\plimsoll}}}{N_{4}} & 0 & 0 & 0 & 0 \\
0 & 0 & 0 & 0 & 0 \\
0 & 0 & 0 & 0 & 0 \\
0 & 0 & 0 & 0 & 0 \\
0 & 0 & 0 & 0 & 0 \\
0 & \frac{2k_{\rm on}e^{\ddagger}}{N_{9}} & 0 & 0 & 0 \\
0 & 0 & \frac{2k_{\rm on}e^{*}}{N_{10}} & 0 & 0 \\
0 & 0 & 0 & \frac{2k_{\rm on}e^{\dagger}}{N_{11}} & 0 \\
0 & 0 & 0 & 0 & \frac{2k_{\rm on}e^{*}}{N_{12}}
\end{array}\right),\]
\end{frame}
\normalsize
where \(N_{x}\) normalises each row across both \(M\) and \(T\) \(x\) to unity. 

The sub matrix \(M\) quantifies the transitions between non-absorbing states. The sub matrix \(T\) quantifies transitions into the absorbing states. The bottom left quadrant is a zero matrix and the bottom right quadrant is the identity matrix representing the probability of leaving the absorbing states. From here we can find the fundamental matrix \(W=(I-M)^{-1}\), which quantifies the number of times the system visits a non-absorbing state on the way to being absorbed, given that it started in a particular state. From here we can calculate \(Z=W.T\). \(Z_{xy}\) gives the probability of eventually reaching absorbing state $12+y$ given a starting state $x$.

The overall rate in steady state at which the system produces dimers with the sequence \(1,1\) (absorbing state \(y=2\)) from monomers is simply \(\Phi_{1,1}^{\rm create} = \pi_0 \sum_{x=1}^{4} k_{0x} Z_{x2}\). Other rates for production, degradation and interconversion of dimers can be calculated similarly, giving
\small
\begin{equation}
    \Phi^{\rm create}_{2,2}/\pi_0 = [2]k_{\rm on}Z_{1,2} +[2]k_{\rm on}Z_{2,2} + 
    [1]k_{\rm on}Z_{3,2} + [1]k_{\rm on}Z_{4,2},\notag
\end{equation}
\begin{equation}
    \Phi^{\rm create}_{1,2}/\pi_0 = [2]k_{\rm on}Z_{1,3} +[2]k_{\rm on}Z_{2,3} + 
    [1]k_{\rm on}Z_{3,3} + [1]k_{\rm on}Z_{4,3},\notag
\end{equation}
\begin{equation}
    \Phi^{\rm create}_{2,1}/\pi_0  = [2]k_{\rm on}Z_{1,4} +[2]k_{\rm on}Z_{2,4} + 
    [1]k_{\rm on}Z_{3,4} + [1]k_{\rm on}Z_{4,4},\notag
\end{equation}
\begin{equation}
    \Phi^{\rm create}_{1,1}/\pi_0  = [2]k_{\rm on}Z_{1,5} +[2]k_{\rm on}Z_{2,5} + 
    [1]k_{\rm on}Z_{3,5} + [1]k_{\rm on}Z_{4,5},
\end{equation}
\footnotesize
\begin{equation}
    \Psi^{\rm destroy}_{2,2}/\pi_0  = 2k_{/\pi_0 \rm on}Z_{9,1},\hspace{5mm}
    \Psi^{\rm destroy}_{2,1}/\pi_0  = 2k_{\rm on}Z_{10,1},\hspace{5mm}
    \Psi^{\rm destroy}_{1,2}/\pi_0  = 2k_{\rm on}Z_{12,1},\hspace{5mm}
    \Psi^{\rm destroy}_{1,1}/\pi_0  = 2k_{\rm on}Z_{11,1},
\end{equation}
\begin{equation}
\Psi^{\rm switch}_{1,1\rightarrow1,2}/\pi_0 =2k_{\rm on}Z_{11,3},\hspace{5mm}
\Psi^{\rm switch}_{1,1\rightarrow2,1}/\pi_0 =2k_{\rm on}Z_{11,5},\hspace{5mm}
\Psi^{\rm switch}_{1,1\rightarrow2,2}/\pi_0 =2k_{\rm on}Z_{11,2},\notag
\end{equation}
\begin{equation}
\Psi^{\rm switch}_{1,2\rightarrow1,1}/\pi_0 =2k_{\rm on}Z_{10,4},\hspace{5mm}
\Psi^{\rm switch}_{1,2\rightarrow2,1}/\pi_0= k_{\rm on}Z_{10,5},\hspace{5mm}
\Psi^{\rm switch}_{1,2\rightarrow2,2}/\pi_0 =2k_{\rm on}Z_{10,2}, \notag
\end{equation}
\begin{equation}
\Psi^{\rm switch}_{2,1\rightarrow1,2}/\pi_0 =2k_{\rm on}Z_{12,3},\hspace{5mm}
\Psi^{\rm switch}_{2,1\rightarrow1,2}/\pi_0 =2k_{\rm on}Z_{12,4},\hspace{5mm}
\Psi^{\rm switch}_{2,1\rightarrow1,2}/\pi_0=2k_{\rm on}Z_{12,2}, \notag
\end{equation}
\begin{equation}
\Psi^{\rm switch}_{2,2\rightarrow1,2}/\pi_0=2k_{\rm on}Z_{9,3},\hspace{5mm}
\Psi^{\rm switch}_{2,2\rightarrow1,2}/\pi_0=2k_{\rm on}Z_{9,5},\hspace{5mm}
\Psi^{\rm switch}_{2,2\rightarrow1,2}/\pi_0=2k_{\rm on}Z_{9,4},
\end{equation}
\normalsize
where \(\Psi_{x,y}\) is a rate per unit concentration of dimer \(x,y\), and \(\Phi\) is an absolute rate.
From here we can identify the total fluxes \(J({\bf s})\):
\footnotesize
\begin{equation}
   J(1,1)= \Phi^{\rm create}_{1,1} + [1,2]\Psi^{\rm switch}_{1,2\rightarrow1,1} + [2,1]\Psi^{\rm switch}_{2,1\rightarrow1,1} + [2,2]\Psi^{\rm switch}_{2,2\rightarrow1,1}- [1,1]\left( \Psi^{\rm destroy}_{1,1} + \Psi^{\rm switch}_{1,1\rightarrow1,2} + \Psi^{\rm switch}_{1,1\rightarrow2,1} + \Psi^{\rm switch}_{1,1\rightarrow2,2}\right),\notag
\end{equation}
\begin{equation}
    J(1,2)=\Phi^{\rm create}_{1,2} + [1,1]\Psi^{\rm switch}_{1,1\rightarrow1,2} + [2,1]\Psi^{\rm switch}_{2,1\rightarrow1,2} + [2,2]\Psi^{\rm switch}_{2,2\rightarrow1,2} - [1,2]\left( \Psi^{\rm destroy}_{1,2} + \Psi^{\rm switch}_{1,2\rightarrow1,1} + \Psi^{\rm switch}_{1,2\rightarrow2,1} + \Psi^{\rm switch}_{1,2\rightarrow2,2}\right),\notag
\end{equation}
\begin{equation}
    J(2,1)=\Phi^{\rm create}_{2,1} + [1,2]\Psi^{\rm switch}_{1,2\rightarrow1,1} + [1,1]\Psi^{\rm switch}_{1,1\rightarrow2,1} + [2,2]\Psi^{\rm switch}_{2,2\rightarrow2,1} -  [2,1]\left( \Psi^{\rm destroy}_{2,1} + \Psi^{\rm switch}_{2,1\rightarrow1,2} + \Psi^{\rm switch}_{2,1\rightarrow1,1} + \Psi^{\rm switch}_{2,1\rightarrow2,2}\right),\notag
\end{equation}
\begin{equation}
    J(2,2)=\Phi^{\rm create}_{2,2} + [2,2]\Psi^{\rm switch}_{2,2\rightarrow1,1} + [2,2]\Psi^{\rm switch}_{2,2\rightarrow1,2} + [2,2]\Psi^{\rm switch}_{2,2\rightarrow2,1} - [2,2]\left( \Psi^{\rm destroy}_{2,2} + \Psi^{\rm switch}_{2,2\rightarrow1,2} + [2,1]\Psi^{\rm switch}_{2,2\rightarrow2,1} + \Psi^{\rm switch}_{2,2\rightarrow1,1}\right).
\end{equation}
\normalsize

\section{Model of copy production for oligomers of length \(|s|>2\)}

The model, schematically illustrated in Fig.~\ref{fig4}\,(a), is adapted from temporary thermodynamic discrimination model in Ref. \cite{Poulton}. We first describe the model of Ref. \cite{Poulton}, steps (2)-(5) in Fig.~\ref{fig4}\,(a), before outlining the extension considered here. 

We consider a copy oligomer \({\bf s}=s_{1},...,s_{l}\), made up of a series of sub-units or monomers \(s_{x}\), growing with respect to a template \({\bf n}=n_{1},...,n_{L}\) (\(l\leq L\)). Inspired by transcription and translation, we consider a copy that detaches from the template as it grows. We consider whole steps in which a single monomer is added or removed, encompassing many individual chemical sub-steps. As illustrated in Figure \ref{fig4}(a), after each step there is only a single inter-polymer bond at position \(l\), between $s_l$ and $n_l$. As a new monomer joins the copy at position \(l+1\), the bond position \(l\) is broken.

As in the dimerisation model considered earlier in the text, we shall consider a template polymer \({\bf n}\) made entirely of monomers of type 1. Given the assumed symmetry between the interactions of the two monomer types, and equal concentrations of the monomer baths as used throughout this work, the results apply equally well to any template sequence. Monomers of type 1 can simply be interpreted as correct matches and monomers of type 2 as incorrect matches for any template sequence \({\bf n}\). 

Having defined the model's state space, we now consider state free energies. By analogy with the dimerisation model, we define \(\Delta G^\plimsoll_{\rm{dim}}\) as the free-energy change of adding a specific monomer to the end of the copy oligomer at standard concentration. The environment contains baths of monomers; a monomer of type \(s\) has a constant concentration \([s]\) relative to the standard concentration. The chemical free-energy change for the transition between any specific sequence \(s_{1},...,s_{l}\) and any specific sequence \(s_{1},...,s_{l+1}\), ignoring any contribution from interactions with the template, is then \(\Delta G^\plimsoll_{\rm{dim}} - \ln [s_{l+1}]\).

We then consider the effect of specific interactions with template. We again define $\Delta G_{r/w}^{\plimsoll}$ as the binding free energies for matched/mismatched monomers and the template. Overall, each forward  step makes and breaks one copy-template bond. There are four possibilities for forward steps:  either adding 1 or 2 at position \(l+1\) to a copy with \(s_l = 1\); or adding 1 or 2 in position \(l+1\) to a copy with \(s_l = 2\). The first and last of these options preserve the same interaction with the template, so the total free-energy change for monomer addition is \(\Delta G^\plimsoll_{\rm{dim}} - \ln[s_{l+1}]\). For the second case a correct bond is broken and an incorrect bond added, implying a  free-energy change of  \(\Delta G^\plimsoll_{\rm{dim}}-\Delta G_{r}^{\plimsoll}+\Delta G_{w}^{\plimsoll}-\ln[2]\).  Conversely, for the third case, an incorrect bond is broken and a correct bond added, giving a free-energy change of \(\Delta G^\plimsoll_{\rm{dim}}+\Delta G_{r}^{\plimsoll}-\Delta G_{w}^{\plimsoll}-\ln[1]\).

These free energies constrain the kinetics of transitions between the various states, but are compatible with a range of kinetic models. In the temporary thermodynamic discrimination model of ref.~\cite{Poulton}, all forwards steps are assumed to occur with the same rate, and sequence-based discrimination occurs in the backwards step. For simplicity, in this case we further assume that each step can be modelled as a single transition with an exponential waiting time, yielding:
\begin{align}
\nu_{1,1}^{+}&=[1]k,\\
\nu_{1,2}^{+}&=[2]k,\\
\nu_{2,1}^{+}&=[1]k,\\
\nu_{2,2}^{+}&=[2]k,\\
\nu_{1,1}^{-}&=ke^{\Delta G_{\rm dim}^\plimsoll},\\
\nu_{1,2}^{-}&=ke^{\Delta G_{dim}^\plimsoll-\Delta G_{r}^{\plimsoll}+\Delta G_{w}^{\plimsoll}},\\
\nu_{2,1}^{-}&=ke^{\Delta G_{\rm dim}^\plimsoll-\Delta G_{w}^{\plimsoll}+\Delta G_{r}^{\plimsoll}},\\
\nu_{2,2}^{-}&=ke^{\Delta G_{\rm dim}^\plimsoll}.
\end{align}\\
Here, \(k\) is a rate constant that sets the overall timescale (we take \(k=1\) in reduced units without loss of generality). \(\nu_{i,j}^{+}\) is the rate for adding a monomer of type \(j\) to a copy with a monomer of type \(i\) at the leading edge, and  \(\nu_{i,j}^{-}\) is the reverse process. 

To allow for initiation and termination of copying, we include two additional transitions. Unbinding transitions, whether of the initial monomer or the final monomer of a complete copy, are parameterised by:
\begin{align}
\nu_{1}^{\rm off}=ke^{\Delta G_{r}^{\plimsoll}},\\
\nu_{2}^{\rm off}=ke^{\Delta G_{w}^{\plimsoll}}.
\end{align}\\

Binding of either a monomer to the initial site, or oligomer to the final site, is assumed to have a rate
\begin{align}
\nu_{s}^{\rm on}=k[s],\\
\nu_{\bf{s}}^{\rm on}=k[\bf{s}].
\end{align}\\

Note that, for simplicity, we ignore the (challenging) question of how partial fragments are prevented from binding to or detaching from the template, or the possibility of multiple copies being bound to the template at once.

We simulate the system repeatedly using a Gillespie simulation \cite{Gillespie}, with the system initiated with either a monomer \(s\) sampled from \(t(s)\) seeded at the first site, or an oligomer \({\bf s}\) sampled from \(p({\bf s})\) 
attached to the final position \(L\) of the template. The simulation is allowed to run, and terminates either when a complete oligomer detaches from the final site of the template, or a single monomer detaches from the first site of the template. 

By running many simulations it is possible to calculate the probability of creating a full length oligomer given that a single monomer binds to the template, \(P^{\rm create}\), and the probability of destroying an oligomer given that one binds to the final site on the template, \(P^{\rm destroy}\). Finally \(P^{\rm transform}=1-P^{\rm destroy}\) is the probability of a full oligomer detaching from the template given that an oligomer previously attached at the final site (including those where attached and detached template are identical). This oligomer will have had some subset of its initial monomers transformed through removal of old and addition of new monomers.
 
We set the oligomer concentration 
\begin{align}
    [S_{\rm tot}]=[M_{\rm tot}]^{n}e^{(\Delta G_{\rm dim}^\plimsoll-F)(n-1)}.
\end{align}
Here \(F\) sets the equilibrium position; for \(F=0\), the equilibrium is \(\Delta G_{\rm dim}^\plimsoll=0\), for \(F=3\), the equilibrium is at \(\Delta G_{\rm dim}^\plimsoll=3\) etc. In our model, \(F=5\).

The rate of oligomer creation per unit time in which the template is in an empty state is given by \(\tilde{k}_{\rm create}=[M_{\rm Tot}]kP^{\rm create}\). The rate of oligomer destruction per unit empty template is \(\tilde{k}_{\rm destroy}=k[S_{\rm tot}]P^{\rm destroy}\). The net flux per empty template, \(\tilde{J}_{\rm tot}=\tilde{k}_{\rm create}-\tilde{k}_{\rm destroy}\), is plotted in Fig. \ref{fig4}(a).

We can also calculate the average fraction of  the monomers incorporated during a creation event that are mismatches with the template, \(\epsilon_{\rm create}\). Similarly, the average fraction of incorrect monomers destroyed during a destruction event, \(\epsilon_{\rm destroy}\), can be extracted from simulations. Finally, \(\epsilon_{\rm transform}\) is defined as the difference in average error density between the sequences at the end and start of a transformation event: \(\epsilon_{\rm transform}=\epsilon_{\rm final}-\epsilon_{\rm initial}\). From these quantities we calculate the overall error rate as the proportion of the net number of monomers added to oligomers that are incorrect matches to the template:
\begin{equation}
    \epsilon= \frac{k_{\rm create}\epsilon_{\rm create} - k_{\rm destroy}\epsilon_{\rm destroy} + k[S_{\rm tot}]P^{\rm transform}\epsilon_{\rm transform}}{\tilde{J}_{\rm tot}},
\end{equation}
which is plotted in Fig. \ref{fig4}(b).

\end{document}